\documentclass[usenatbib]{mn2e}
\def\vdot{\ensuremath{\!\cdot\!}}
\usepackage[T1]{fontenc} 
\usepackage{aecompl} 
\usepackage{lachaume}
\usepackage{subfigure}
\usepackage{relsize}
\usepackage{notations}
\graphicspath{{./pdf/}}
\def\conj#1{{#1}^*}
\let\wideexp=\exp
\def\exp#1{\ensuremath{\text{e}^{#1}}}

\def\Nu{\scalebox{1.6}{\ensuremath{\nu}}}

\renewcommand\paragraph[1]{\vskip1em\noindent\textit{#1}}
\usepackage{times}
\begin{document}

\title[Bandwidth smearing in IR interferometry]{Bandwidth smearing in infrared long-baseline interferometry.\\
       Application to stellar companion search in fringe-scanning mode.}
\author[R. Lachaume \& J.-P. Berger]
  {R.~Lachaume$^{1,2}$ \& J.-P.~Berger$^{3}$\\
     $^1$ Centro de Astroingenier\'\i a, 
     Instituto de Astrof\'\i sica, Facultad de F\'\i sica, 
     Pontificia Universidad Cat\'olica de Chile,
     Casilla 306, Santiago 22, Chile\\
     $^2$ Max-Planck-Institut f\"ur Astronomie,
     K\"onigstuhl 17,
     D-69117 Heidelberg\\
     $^3$ European Southern Observatory}

\maketitle

\begin{abstract}
{In long-baseline interferometry, bandwidth smearing of an extended source occurs \rev{at} finite bandwidth when its different components produce interference packets that only partially overlap.  In this case, traditional model fitting or image reconstruction using standard formulas and tools lead to biased results.  
 
  In this paper, we propose and implement a method to overcome this effect by calculating analytically a corrective term for the \rev{conventional} interferometric observables: the visibility amplitude and closure phase.  For that purpose, we model the interferogram taking into account the finite bandwidth and the instrumental differential phase.  We obtain generic expressions for the visibility and closure phase in the case of temporally-modulated interferograms, either processed using Fourier analysis or with the ABCD method. The expressions can be used to fit arbitrary models to the data. We then apply our results to the search and characterisation of stellar companions with PIONIER at the Very Large Telescope Interferometer, assessing the bias on observables and model-fitted parameters of a binary star. Finally, we consider the role of the atmosphere, first with an analytic model to identify the main contributions to bias and\rev{, secondly, by confirming the model} with a numerical simulation of the atmospheric turbulence. 

  In addition to \rev{the analytic expressions}, the main results of our study are: \rev{(i)} the chromatic dispersion in the beam transport in the instrument has a strong impact on the closure phase and \rev{introduces} additional biases even at separations where smearing is not expected to play an important role; \rev{(ii)} the atmospheric turbulence introduces additional biases when smearing is present, but the impact is important only at very low spectral resolution; \rev{(iii)} the bias on the observables strongly depends on the recombination scheme and data processing; \rev{(iv)} the goodness of model fits \rev{is improved} by modelling a Gaussian bandpass as long as the smearing is moderate.
}

\end{abstract}

\section{Introduction}
\begin{figure*}
\includegraphics{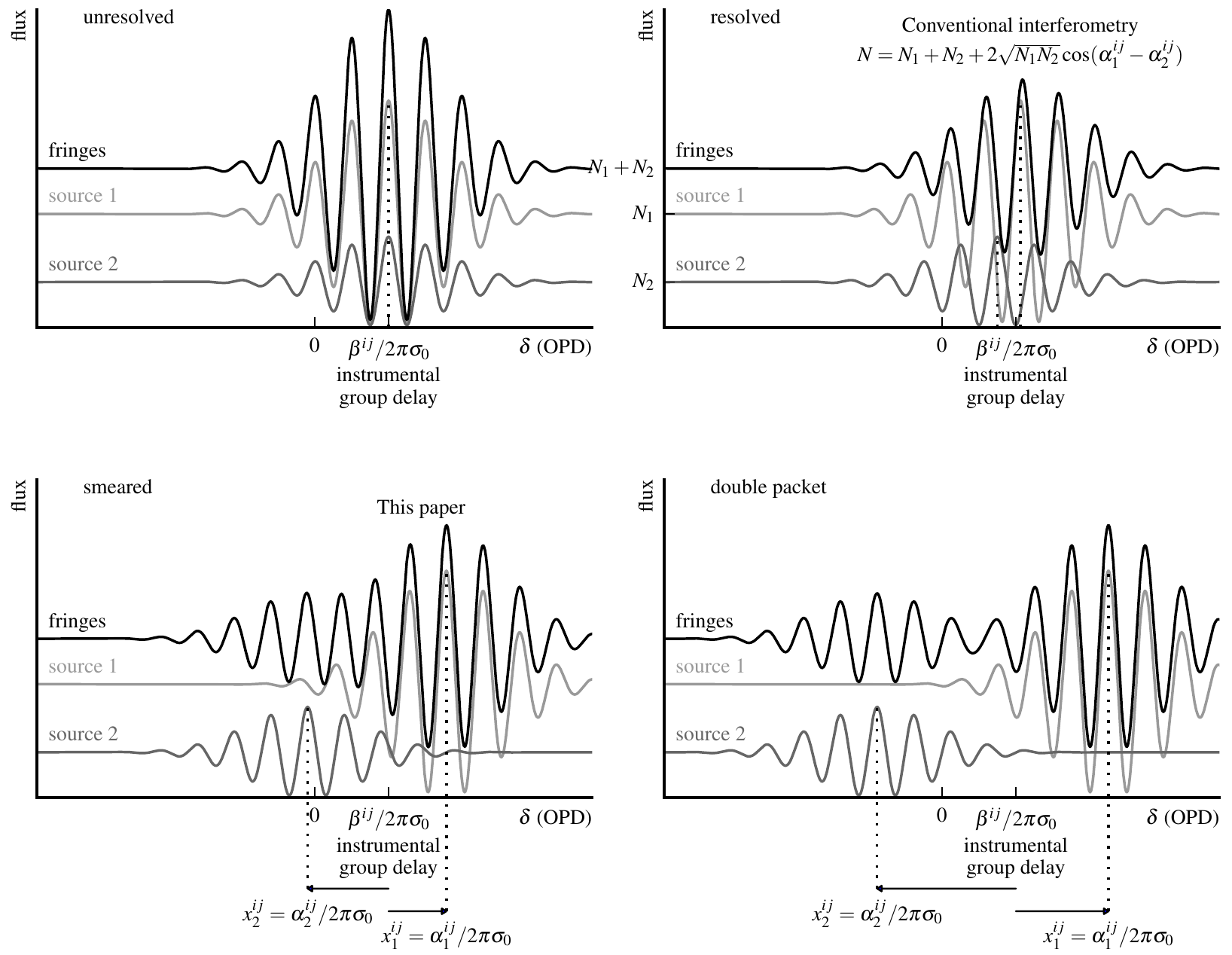}
\caption{The transition to the double fringes packet for a binary. \rev{Gray} curves show the fringes of the individual components \rev{with vertical dotted lines indicating the position of their centres}. The \rev{black} curve is the detected fringe packet (sum). In comic strip order: (i) An unresolved binary superimposes two  fringe systems and achieve maximum fringe contrast; (ii) Contrast loss arises when the binary is resolved because individual fringe packets do not overlap exactly; (iii) The packet is elongated and loses its original shape when the binary is sufficiently separated, this is the transition to (iv) two separate fringe packets are clearly seen.  Cases (i), (ii) are standard in interferometry. Case (iv) is easily analysed, but helps to understand why smearing occurs: each fringe packet doesn't seem impacted, but the power in its fringes is diluted by the incoherent flux from the other one; the resulting visibility will be a constant, strictly smaller than one, independent \rev{of} binary separation.  This paper focuses on case (iii) that has not been thoroughly studied in the optical. Some of the notations of the paper are also reported: $\beta^{ij}$, the decentering of the fringe packet of an on-axis source due to instrumental effects; $\xobj[ij]{o}$ the OPD position shift of the fringe packet for an off-axis source; $\phiobj[ij]{o}$ the same as the latter expressed in terms of a phase shift.}
\label{fig:smearing-explanation}
\end{figure*}

Long-baseline interferometry is an observational technique used from the optical \citep{MIC20} to the radio domain \citep{RYL46,PAW46} that allows to overcome the resolution limit of single-dish telescopes, as ultimately set by diffraction. To achieve such a goal an ideal interferometer measures the complex degree of coherence and relates this so-called complex visibility to the object intensity distribution through a Fourier transform \citep{VCI34,ZER38}.  Practically speaking, interference fringes are formed and their contrast and shift will be used to retrieve partial or total information on the complex visibilities.

There are numerous sources of error and biases that have to be evaluated and as much as possible corrected in order to provide a proper estimation of the interferometric observables. Among them, \emph{bandwidth smearing} occurs in finite bandwidth for \rev{objects spanning an extended field of view}. The interferogram corresponding to a point-like source has a coherence length of the order of $R\lambda$ where $R$ is the spectral resolution and $\lambda$ the central wavelength.  For two points of an extended sourc\rev{e} separated by a distance $\theta$ along the projected baseline length $B$ corresponding to two telescopes of the array, individual fringe packets are shifted with \rev{respect to} each other by an optical path difference $\theta/B$.  When the OPD shift $\theta/B$ becomes of the order of, or greater \rev{than}, the fringe packet width, i.e. when $\theta \approx R\lambda/B$, the fringe packets of these points do not overlap correctly and bandwidth smearing of the interferogram occurs (see bottom left panel of Fig.~\ref{fig:smearing-explanation}).  In other words, one can consider that the coherence length of an interferogram $R\lambda$ corresponds to an angular extension on the sky $\theta \approx  R\lambda/B$: it is called the interferometric field of view.  Objects composed of multiple \rev{incoherent} sources, either discrete or continuous, are affected by the smearing when their extent becomes of the order of the interferometric field of view.

Figure \ref{fig:smearing-explanation} shows an illustration of that effect applied to the case of a binary system. Each of the sources \rev{contributes} with a fringe packet; the observed interferogram is their sum. The distance between the interferograms is proportional to the angular separation. We can distinguish four separation regimes: 1) the unresolved case; 2) the resolved case where the separation is a small fraction of the interferogram envelope; 3) the smeared regime where \rev{the separation} is not anymore a small fraction and interferometric estimators are altered; 4) the ``double packet'' regime where two fringes packets are well separated.

While this effect \rev{has been} known for decades \citep{THO73}, it cannot be remedied by calibration as other biases. This was analysed in a review by \citet{BRI89} in the radio-astronomy context, in which the observer had no other choice but to define, somewhat heuristically, the best compromise between observing performance and limiting the bandwidth smearing. However, modern radio techniques, using \textit{a posteriori} software recombination, can overcome the problem in many situations by using several phase centres, around which smearing does not occur. In the optical and the infrared, software recombination is not technically feasible and bandwidth smearing must be dealt with. \citet{ZHA07} \rev{recommend to limit} the field of view $\theta$ to $1/5$ of \rev{the} theoretical value \rev{of the interferometric field of view} i.e. $\theta\approx R\lambda/(5B)$ to remain in the resolved regime. For an interferometer working in the near-IR with 100\,m baselines, this corresponds to 5--10 milliarcseconds of separation when a \rev{wide-band filter ($\lambda/\Delta\lambda \sim \rev{5}$)} is used without spectral resolution.  The main leverage to increase the interferometric field of view is adapting the spectral resolution or the baseline length. However, it comes very often at a prohibitive sensitivity cost (spectral resolution) or a loss of spatial \rev{resolution} (baseline length).

In this paper, we present the first analytic calculation of the bandwidth smearing effect on the two main optical interferometric observables, namely the squared visibility and closure phase. We restricted the calculation to temporally encoded interferograms, including the so-called Fourier mode \rev{(a full scan of the fringe packet)} and the \rev{temporal} ABCD \rev{(a 4-point scan of a central fringe)}, which are among the most popular optical schemes. Fourier mode has been or is being used at COAST \citep{COAST}, IOTA with IONIC \citep{IONIC} and FLUOR \citep{IOTAFLUOR}, CHARA with FLUOR \citep{CHARAFLUOR} and CLIMB \citep{CLIMB}, VLTI with VINCI \citep{VINCI}, PIONIER \citep{PIONIER2,PIONIER}, and MIDI \citep{MIDI}. \rev{Temporal} ABCD is the choice at PTI \citep{PTI} and the Keck Interferometer \citep{KI}.  It should be stressed that a similar line of reasoning can be used with very little adaptation to the 8-point time-encoded interferograms of NPOI \citep{NPOI}, and, with more efforts, to spatially encoded interferograms such as in VLTI/AMBER \citep{AMBER} and static ABCD systems such as VLTI/PRIMA \citep{PRIMA}.
The derived formula\newrev{e} can \rev{be} applied to correct \emph{any} squared visibility and closure phase analytic formula describing the object angular intensity distribution. We apply this corrective technique to the study of binary stellar systems. Indeed optical long baseline interferometry is a unique tool to study the astrometry of close binary systems with milli-arcsecond accuracy to provide direct means to measure accurate masses. Moreover very recently several attempts at searching for substellar companions \citep{ABS11,ZHA11} are pushing the technique down to dynamical ranges where no adverse effects can be neglected. Since \rev{most studies forgo} bandwidth smearing correction \rev{without assessing the biases that may arise from such approximation}, we felt a proper treatment had become mandatory and would be useful in the future. For practical purposes we used the PIONIER instrument characteristics to provide an application of that work. PIONIER is currently being used at the Very Large Telescope Interferometer \citep[VLTI,][]{VLTI} to combine four beams in the H band ($1.5$ to $1.8\mu\mathrm{m}$).

Sect.~\ref{sec:hypnot} gives the analytic expression of the observables in the absence of atmospheric turbulence for an instrument working in fringe-scanning mode.  Section~\ref{sec:bin} is an application of these formulas to a binary star, which allows us to analyse the bias that smearing produces on the interferometeric observables and the model-fitted parameters of the binary.  We also show there how simulated fringes of PIONIER are much better fitted with the smeared model than with the standard expression. Finally, section~\ref{sec:atm} studies the impact of atmospheric turbulence on the observables, indicating that a moderate spectral resolution is enough to alleviate most of its effects.

\section{Modelling the bandwidth smearing: turbulence-free case}
\label{sec:hypnot}
\label{sec:ana}

%%%% DEFINITIONS  %%%%
\begin{figure*}
\centering
\includegraphics[width=\textwidth]{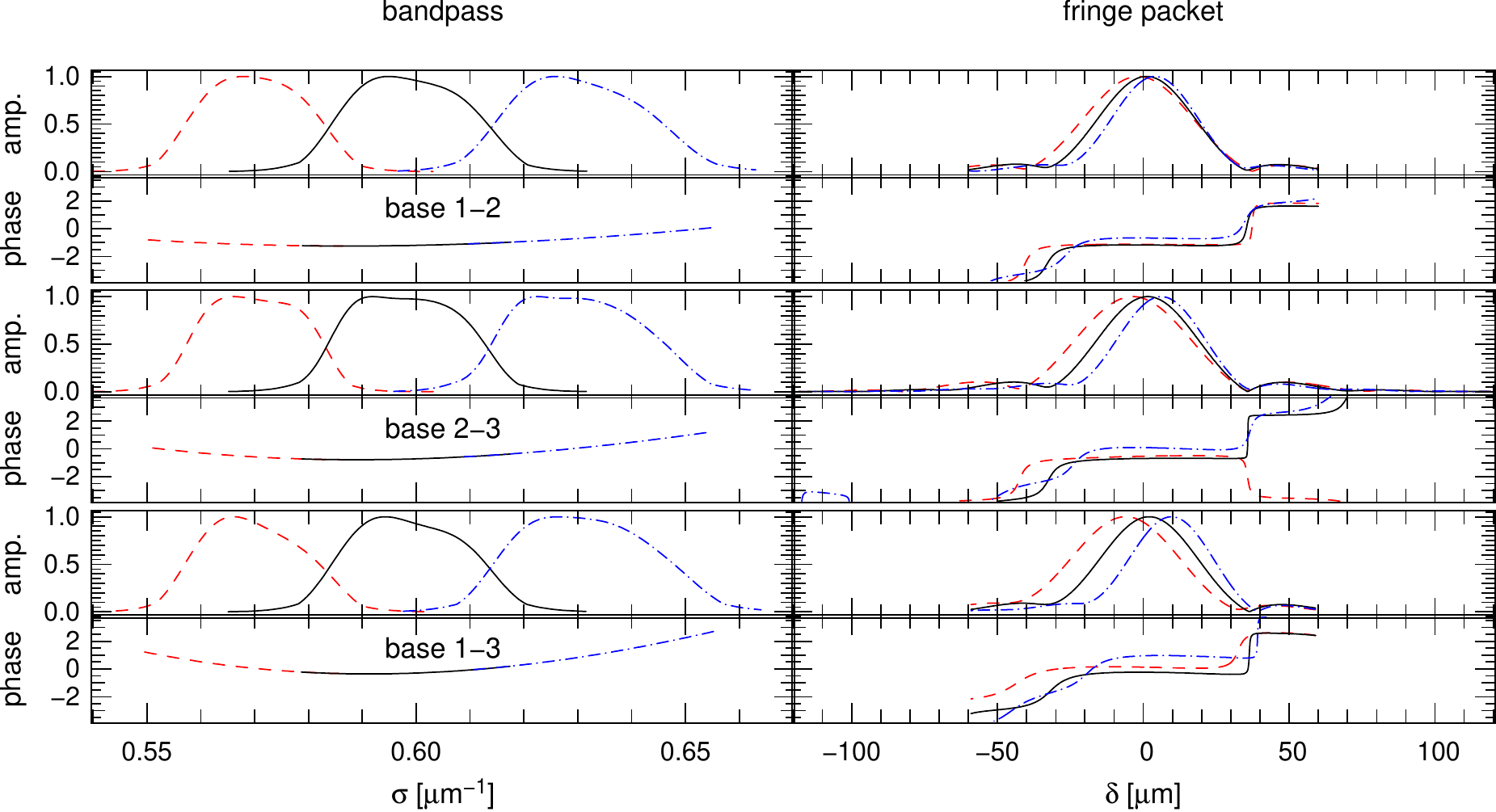}
\caption{Spectral transmission and fringe packet envelope for PIONIER, as measured on an internal calibration with the 3-channel spectral setting of the H band. The left column display\newrev{s} the spectral transmission and instrumental phase for a telescope triplet, as contained in $\textsflux[ij]{\text{lamp}}(\wavenum - \wavenumzero)$. The right column shows the envelope and phase of the fringe packet, given by the Fourier transform of the latter. The slope of the instrumental phase translates into a fringe packet decentering, known as group delay.  \rev{Phases are expressed in radians.}}
\label{fig:PT}
\end{figure*}

\begin{table}
\caption{Principal notations of this paper.}
\label{tab:notations}
\begin{tabular}{ll}
  \hline\hline
  \multicolumn{2}{l}{Indexing}\\
  $o$, $p$, $q$                      & Source number (index)\\
  $i$, $j$, $k$                      & Station number (superscript)\\
  \hline
  \multicolumn{2}{l}{Space and spatial frequency variables}\\
  $\wavenum$, $\wavenumzero$              & Wavenumber, central wavenumber\\
  $\xi = \wavenum - \wavenumzero$         & Reduced wavenumber\\
  $\opdvar$, $\textopd[ij]$               & Optical path difference\\
  \textxobj[ij]{o}                        & Fringe packet position in a perfect instrument\\
  $\textphiobj[ij]{o} = 2\pi\wavenumzero\textxobj[ij]{o}$
  & Fringe packet phase in perfect instrument\\
  $\textphishift[ij]{}$                   & Instrumental group delay (see Fig.~\ref{fig:smearing-explanation})\\
                                          & $\rightarrow$ Packet position is \smash{$\xobj[ij]{o} + \phishift[ij]{}/2\pi\wavenumzero$}\\
  \hline
  \multicolumn{2}{l}{Functions of wavenumber $\wavenum$ or reduced wavenumber $\xi$}\\
  $\sfluxstar{o}(\wavenum)$               & Spectrum of a point source\\
  $\texttrans[i]{}(\xi)$                  & Transmission through an arm\\
  $\loss[ij]{}(\xi)$                      & Instrumental contrast\\
  $\insphi[ij]{}(\xi)$                    & Instrumental differential phase\\
  $\textsflux[ij]{o}(\xi)$                & The equivalent of $\newrev{N_iN_j}$\\
  \hline
  \multicolumn{2}{l}{Functions of OPD $\opdvar$ or phase $\alpha = 2\pi\wavenumzero\opdvar$}\\
  $\textphasor[ij]{}(\opdvar)$            & Complex coherent flux\\
  $\textsmearing{}(\alpha)$               & Complex smearing\\
  $\textenvband{}(\alpha)$                & Smearing amplitude\\
  $\phiband{}(\alpha)$                    & Smearing phase\\
\hline
  \multicolumn{2}{l}{Fluxes}\\
  $\textflux{o}$                          & Flux of a point source\\
  $\textflux{}$                           & Total flux\\
  $\textflux[ij]{op}$                     & Flux product equivalent\\
\hline
  \multicolumn{2}{l}{Other}\\
  $\dopd[ij]{}$                           & OPD scanning speed\\
\hline
\end{tabular}
\end{table}

In order to introduce the basic concepts of the data processing for fringe-scanning mode instruments, we remind \rev{the reader} here how observables are derived \rev{in monochromatic light}.

Ignoring the atmosphere and instrumental signatures, the interferogram of a binary on baseline $ij$ can be written as
\begin{equation}
  N^{ij}(\opdvar) = \rev{N}_1 \big[1 +  \cos 2\pi\wavenum\opdvar\big] 
  + \rev{N}_2\big[ 1 + \cos(2\pi\wavenum\opdvar + \phiobj[ij]{}) \big]\newrev{,}
\end{equation}
where $\rev{N}_1$ and $\rev{N}_2$ are the fluxes of each component, $\opdvar$ is the OPD \rev{between the arms of the interferometer}, and $\textphiobj[ij]{} = (2\pi\wavenum\textbase[ij]\cdot\textpos{})$ is proportional to the binary separation $\textpos{}$, the projected baseline $\textbase[ij]$, and wavenumber $\wavenum$.

It is convenient to use the coherent flux, a complex quantity representing the interferogram, from which the continuum $\rev{N}_1 + \rev{N}_2$ is removed and the negative frequencies are filtered out. In practice, one can take the Fourier transform of the interferogram, remove all frequencies but a small interval centred on the frequency of the fringes, and take the inverse Fourier transform. \rev{The coherent flux can be written as}
\begin{equation}
  \phasor[ij]{}(\opdvar) =  \exp{2\pi\j\wavenum\opdvar} 
  \big[ \rev{N}_1 + \rev{N}_2\, \exp{\j\phiobj[ij]{}} \big].
  \label{eq:Mmono}
\end{equation}

The \rev{square visibility amplitude} is obtained by dividing the power contained in the coherent flux by that in the continuum:  
\begin{equation}
\begin{split}
  \vsqPS[ij]{} &= \frac{<|\phasor[ij]{}(\opdvar)|^2>_\opdvar}{(\rev{N}_1+\rev{N}_2)^2}\\
               &= 1 - \frac{4 \rev{N}_1 \rev{N}_2}{(\rev{N}_1+\rev{N}_2)^2} \sin^2 \frac{\phiobj[ij]{}}2,
\end{split}
  \label{eq:Vmono}
\end{equation}
\rev{where $<x>_\opdvar$ means the average of variable $x$ over the OPD.} In practice, the power may be computed using the Fourier \rev{transform} of the coherent flux, which is strictly equivalent (Parseval's identity).

When a triplet of telescopes $ijk$ is used, the closure phase is used to obtain partial information on the phase because it is independent \rev{of} atmospheric turbulence.  It is the argument of the bispectrum given by:
\begin{equation}
\begin{split}
  \bisp[ijk]{} &=\ <\phasor[ij]{}(\opd[ij](t))
                  \phasor[jk]{}(\opd[jk](t))
                  \phasor[ki]{}(\opd[ki](t))>_\opdvar\\
             &= (N_1-N_2)^2 
                  + 4\rev{N}_1 \rev{N}_2 % \prod_{b \in \{ij,jk,ki\}} 
                         \cos\frac{\phiobj[ij]{}}2 
                         \cos\frac{\phiobj[jk]{}}2 
                         \cos\frac{\phiobj[ki]{}}2 
                    \\
                    &\quad -4\j \, \rev{N}_1 \rev{N}_2(\rev{N}_1-\rev{N}_2) 
                    \sin\frac{\phiobj[ij]{}}2 
                    \sin\frac{\phiobj[jk]{}}2
                    \sin\frac{\phiobj[ki]{}}2 ,
\end{split}
\label{eq:Bmono}
\end{equation}
where $\textopd[ij]$, $\textopd[jk]$, and $\textopd[ki]$ are
the time-modulated OPDs on the three baselines, meeting the closure
relation  $\textopd[ij] + \textopd[jk] + \textopd[ki] = 0$. (Eq.~\ref{eq:Bmono} gives \rev{a compact, generic expression for the bispectrum in the same way \citet{LEB12} did for the specific case of high-contrast binaries.})

The goal of this section is to describe the coherent flux, squared visibility, and closure phase of time encoded interferograms processed by means of Fourier analysis, when observing a source of arbitrary geometry in finite bandwidth. In other words, we seek to generalise Eqs.~(\ref{eq:Mmono}, \ref{eq:Vmono},~\& \ref{eq:Bmono}) and provide ready-to-use formulas to fit object models to smeared data.  For the sake of simplicity we use a discrete formalism valid for a collection of point-like sources.   The results presented here are easily generalised to systems of resolved, compact sources (Appendix~\ref{ap:syscomp}) and to any system with our summations over a finite number of point-like sources replaced by integrations on the plane of the sky.

The most \rev{frequently used} notations and symbols used in this section are given in Table~\ref{tab:notations}.

\subsection{Interferogram}
\label{sec:interferogram}

We consider an interferometer with stations $i$, $j$, etc. separated  by a baseline $\base[ij]$ operating in a spectral channel centred on wavenumber $\wavenumzero$.  In the following developments we shall use $\wavenum$, the wavenumber, and $\xi = \wavenum - \wavenumzero$ as ``reduced'' wavenumber.  Without losing generality, we assume that we observe an object made of several point sources $o$, $p$, etc. with positions $\textpos{o}$, $\textpos{p}$, etc. \rev{in} the plane of the sky and spectra $\textsfluxstar{o}(\wavenum)$, $\textsfluxstar{p}(\wavenum)$, etc.

The interferometer measures the complex coherent flux of the electromagnetic field by forming dispersed fringes on a detector.  In our case, fringes are obtained by a temporal modulation of the optical path difference (OPD) $\opdvar$ around an ideal position $\xobj[ij]{o}$. This position is related to the angular position of the source in the sky $\pos{o}$ through the relation $\xobj[ij]{o} = \base[ij]\vdot\pos{o}$. Each of the point sources contributes to a quasi-monochromatic interferogram per instrument spectral channel. Once the incoherent photometric contribution has been removed from the two telescopes and the negative frequencies have been filtered out in Fourier space, the  complex coherent flux of one source reads: 

\begin{equation}
  \phasor[ij]{o}(\xi,\opdvar) = 
               2\sflux[ij]{o} (\xi)
              \exp{
                2\jpi(\wavenumzero+\xi)(\xobj[ij]{o} + \opdvar) 
              }
  \label{eq:phasormono}
\end{equation}

where $\sflux[ij]{o} (\xi)$ is the \rev{``instrumental'' coherent flux density} \rev{primarily} due to the wavelength-dependent instrumental effects\rev{, but also to some extent to the spectrum of the source.} We can define this coherent flux density as:

\begin{equation}
\sflux[ij]{o}(\xi) = \loss[ij]{}(\xi)\sqrt{\trans[i]{}(\xi)\trans[j]{}(\xi)}
                    \,\exp{\j \insphi[ij]{}(\xi)}
                    \,\sfluxstar{o}(\wavenumzero + \xi)
      \label{eq:cohernorm}
\end{equation}
where:
\begin{itemize}
  \item $\loss[ij]{}(\xi)$, is the instrumental visibility, or instrumental contrast loss, \newrev{and} has different origins such as differential polarisation or wavefront aberrations; 
  \item $\insphi[ij]{}(\xi)$, is the instrumental differential phase, \newrev{and} arises from a difference of optical path lengths between the arms of the interferometer that depends on the wavelength. For example this can be the case when light travels through glass (e.g waveguides, dichroics) that do not have the same refraction index dependence as a function of wavelength.
  \item $\trans[i]{}(\xi)$, is the spectral transmission through an arm including detector efficiency.
\end{itemize}
We assume that these instrumental signatures do not depend on the \newrev{OPD position in the interferogram}, which is a good approximation in fringe-scanning mode\newrev{, since the OPD modulation is obtained through a few micrometres of air or vacuum, with negligible dispersion. In other words, we assume that the instrumental differential phase is a static term that is not impacted by the movement of the differential delay lines.} However, this is usually not true for spatially dispersed fringes \citep[see][for a generic expression for the fringes]{TAT06}, so that our approach needs adaptation to instruments like AMBER \citep{AMBER}.

It is now possible to describe the coherent flux for an arbitrary number of sources and across a wider spectral bandpass:

\begin{equation}
  \phasor[ij]{}(\opdvar) = 
     \intinf \sum_o \phasor[ij]{o}(\xi, \opdvar) \idiff\xi, 
  \label{eq:phasorgen}
\end{equation}

For practical purposes we use the Fourier transform
\begin{equation}
  \IFT{f}(\opdvar) =  \intinf f(\xi) \exp{2i\pi\xi\opdvar} \idiff\xi,
\end{equation}
substitute Eq.~(\ref{eq:phasormono}) into Eq.~(\ref{eq:phasorgen}), and
obtain 
\begin{align}
  \phasor[ij]{} (\opdvar) = 
         \sum_o
           2\IFT{
             \sflux[ij]{o}
            }(\xobj[ij]{o} + \opdvar)
            \,\exp{2\jpi\wavenumzero\opdvar + \j\phiobj[ij]{o}}.
  \label{eq:def:phasor}
\end{align}
where $\textphiobj[ij]{o} = 2\pi\wavenumzero\textxobj[ij]{o}$. In the following, we will use the coherent flux expression (Eq.~\ref{eq:def:phasor}) to compute the most \rev{commonly used} interferometric observables i.e. the square visibility and the closure phase.  In practice, $\textsflux[ij]{o}$ is not known a priori. However, it can be inferred from fringes obtained on an internal lamp.  The coherent flux of the lamp fringes yield $\textsflux[ij]{\text{lamp}}$ (see Eq.~\ref{eq:def:phasor}).  \rev{If both the spectrum of the source $\textsflux[\star]{o}$ and that of lamp $\textsflux{\text{lamp}}$ are known, $\textsflux[ij]{o} = \textsflux[ij]{\text{lamp}} \textstrans[ij]{\text{int}} \, (\textsflux[\star]{o}/\textsflux{\text{lamp}})$ (see Eq.~\ref{eq:cohernorm}) where $\textstrans[ij]{\text{int}}$ is the transmission of the interferometer before the calibration lamp.  The amplitude of the VLTI transmission is a smooth function of wavelength that can be considered constant.  Its phase results from dispersive elements in the optical path.  The optical elements of the VLTI before PIONIER are all in reflection and the most dispersive ones (the M9 dichroics) have been designed to display the least differential dispersion, so that the dispersion is dominated by the air in the non evacuated delay line. In the rest of this paper, we have considered near-zenithal observations for which the interferometric delay is small so that the air dispersion could be ignored as Appendix~\ref{ap:gd} shows. While the presence of dispersion in non zenithal observations has a significant impact on the amount of smearing, it neither changes its order of magnitude nor the general conclusions of this paper. When the atmospheric dispersion must be tackled, it can be done either explicitly (Appendix~\ref{ap:gd} explains how) or implicitly by letting the parameters of Sect.~\ref{sec:isr} free in model fits, as \citet{ZHA07} do for the spectral resolution.}

As an illustration, we show in the left panels of Fig.~\ref{fig:PT} the spectral coherence transmission \textsflux[ij]{\text{lamp}} (amplitude and phase) that we measured on the internal source of PIONIER using three spectral channels across the H band on three baselines.  The right panels correspond to the coherent flux of the fringes \textphasor[ij]{\text{lamp}} (amplitude and phase). 

%%% SPECTRAL TRANSMISSION %%%

\subsection{Instrumental spectral response}
\label{sec:isr}
In this paper, after providing generic formulas using Fourier formalism, we will also give closed form expressions for direct use.  To do so, we need an analytic description of the instrumental transmission ($\texttrans[i]{}$) and differential phase ($\textinsphi[ij]{}$).  PIONIER's \rev{instrumental coherent flux density} is obtained on a calibration lamp (Fig.~\ref{fig:PT}, left panels)\newrev{. It} displays a near-quadratic behaviour of the differential phase and a spectral transmission intermediate between top-hat and Gaussian functions.

We therefore describe the instrumental differential phase as
\begin{equation}
  \insphi[ij]{} (\xi) = \insphi[ij]{}(0) + \phishift[ij]{} (\xi/\wavenumzero) + \disp[ij]{} (\xi/\wavenumzero)^2. 
  \label{eq:hyp:insphi}
\end{equation}

The linear term $\textphishift[ij]{}$ in the instrumental differential phase $\textinsphi[ij]{}(\xi)$ translates into a fringe packet shift of $\textphishift[ij]{}/2\pi\wavenumzero$ with respect to the nominal zero OPD (see Fig.~\ref{fig:smearing-explanation}, bottom right panel).  It is called group delay.  In a single-spectral channel interferometer it is possible to zero it by means of fringe tracking. When several spectral channels are observed at the same time, it is no longer possible to do so in all channels simultaneously. \rev{For instance, if a central \rev{spectral} channel is centred at zero OPD, adjacent channels may be shifted with respect to it if there is a differential behaviour of the dispersive elements (such as waveguides, dichroics, or air whose refractive index depend on wavelength) in the beam paths before the recombiner. In the bottom panels of Fig.~\ref{fig:PT} (baseline 1-3), the central \rev{spectral} channel is approximately centred at zero OPD (the solid line on the right panel \newrev{shows the envelope of the fringe packet, i.e. the amplitude of the coherent flux}) with a slope of the phase averaging to $\approx 0$ (same line of the left panel). The adjacent channels feature some shift (dashed lines on the right panels) and non-zero phase slope (same lines on the left). Appendix~\ref{ap:gd} gives a further description of the group delay and its correction through fringe-tracking.} 

The quadratic term in the instrumental differential phase $\disp[ij]{}$ has a less visible impact on the fringe packet.

We will give results both for Gaussian and top-hat transmissions of FWHM $\dwavenum{}$:
\begin{align}
  \transG[i]{}(\xi)  &= \wideexp{-\frac{4 \log 2}{\dwavenum{}^2} \xi^2},
    \label{eq:hyp:bandpass}\\
  \transH[i]{}(\xi)  &=
    \begin{cases}
      1 \quad &\text{if $|\xi| \le \dwavenum{}/2$},\\
      0 \quad &\text{otherwise}.
    \end{cases}
\end{align}

  %%%% VISIBILITY %%%%

\subsection{Square visibility amplitude}
The square visibility amplitude is obtained from the coherent flux using:
\begin{equation}
   \vsqPS[ij]{} 
      = \frac1{4\normtot[ij]{}}
       \intinf \phasor[ij]{}(\opdvar)
            \!\cdot\! \conj{\phasor[ij]{}(\opdvar)} \idiff\opdvar,
   \label{eq:def:vsqPS}
\end{equation}
where \textnormtot[ij]{} is a normalisation factor that relates to the total flux of the target ($\propto \textfluxtot{}^2$) and \rev{$\conj{x}$ stands for the complex conjugate of $x$}. In the first line of the previous equation, we substitute Eq.~(\ref{eq:def:phasor}) and expand the product into a double sum to find\rev{:}
\begin{equation}
  \vsqPS[ij]{}
  = \frac1{\normtot[ij]{}} \sum_{o,p} 
    \exp{\j(\phiobj[ij]{o} - \phiobj[ij]{p})}
    \intinf
      \IFT{\sflux[ij]{o}}(\xobj[ij]{o} + \opdvar) 
      \IFT{\sflux[ij]{p}}(-\xobj[ij]{p} - \opdvar) 
      \idiff\opdvar.
\end{equation}
Using the change of variables $\opdvar \rightarrow u = \opdvar + \textxobj[ij]{o}$, a correlation of Fourier transforms is recognised and simplified into the Fourier transform of a product.  Thus,
\begin{equation}
  \vsqPS[ij]{} = \frac1{\normtot[ij]{}} \sum_{o, p} 
      \IFT{\sflux[ij]{o}\sflux[ji]{p}}(\xobj[ij]{o} - \xobj[ij]{p})
                 \exp{\j(\phiobj[ij]{o} - \phiobj[ij]{p})}.
\end{equation}

The bandwidth smearing is contained in $\IFT{\sflux[ij]{o}\sflux[ji]{p}}$.  It
can be made clearer by introducing the complex smearing
\begin{equation}
  \smearing[ij]{op}(\alpha) = \frac{
     \IFT{\sflux[ij]{o}\sflux[ji]{p}}(\alpha / 2\pi\wavenumzero) 
            }{  \IFT{\sflux[ij]{o}\sflux[ji]{p}}(0)},
  \label{eq:gen:V2smearing}
\end{equation}
\rev{where $\alpha$ is an angular variable that is linked to the OPD by the relation $\alpha = 2\pi\wavenumzero\delta$.} It \rev{is} convenient to use the amplitude and phase \rev{of the smearing}: $\textenvband[ij]{op} = |\textsmearing[ij]{op}|$ is the contrast loss due to smearing and $\textphiband[ij]{op} = \arg \textsmearing[ij]{op}$ is a phase shift induced by it.  We also define the flux product equivalent---the equivalent to $\flux{o}\flux{p}$ in the monochromatic case---as
\begin{equation}
  \norm[ij]{op} = \intinf \sflux[ij]{o}(\xi)\sflux[ji]{p}(\xi)\idiff\xi.
  \label{eq:def:norm}
\end{equation}
With these definitions, we can rearrange the square visibility amplitude:
\begin{equation}
\begin{split}
   \vsqPS[ij]{} =
   \sum_o & \frac{\norm[ij]{oo}}{\normtot[ij]{}} 
        + \sum_{o < p}
        \Bigg[\frac{2\norm[ij]{op}}{\normtot[ij]{}}
        \envband[ij]{op}(\phiobj[ij]{o}-\phiobj[ij]{p})\\
        &\times \cos \left(\phiobj[ij]{o}-\phiobj[ij]{p} 
        + \phiband[ij]{op}(\phiobj[ij]{o}-\phiobj[ij]{p})
           \right) \Bigg]. 
\end{split}
\label{eq:gen:vsqPS}
\end{equation}
These results are independent of the instrumental phase $\insphi[ij]{}$.  If $\textenvband[ij]{op} = 1$ and $\textphiband[ij]{op} = 0$ (no smearing) this formula is equivalent to the monochromatic case (Eq.~\ref{eq:Vmono} in the case of a binary). In practice, model-fitting of square visibility amplitudes by multiple stellar systems uses Eqs.~(\ref{eq:gen:V2smearing}, \ref{eq:def:norm},~\& \ref{eq:gen:vsqPS}).  Knowledge of $\textsflux[ij]{o}$, needed in Eqs.~(\ref{eq:gen:V2smearing} \& \ref{eq:def:norm}), can be inferred from fringes obtained on a calibration lamp (or a calibrator) if the spectra of both lamp and source $o$ are known, as we discussed in Sect.~\ref{sec:interferogram}. 

\def\Vins{\ensuremath{V_\text{ins}}}
When the different sources share the same spectrum, i.e. $\sfluxstar{o}(\xi) \propto \sfluxstar{p}(\xi)$, we may express the visibility as a function of the individual fluxes \textflux{o} and the total flux \textfluxtot{}. In Eq.~\ref{eq:gen:vsqPS}, we then use the flux products in lieu of the flux products equivalents, i.e. $\textflux[ij]{op} = \Vins\textflux{o}\textflux{p}$ and $\textflux[ij]{} = \textflux{}^2$, where 
\begin{equation}
  \Vins^2 = \intinf \loss[ij]{}(\xi)^2\trans[i]{}(\xi)\trans[j]{}(\xi)
  \sfluxstar{}(\xi)^2 \idiff\xi\, \Big/ \intinf \sfluxstar{}(\xi)^2 \idiff\xi 
\end{equation}
is the ``instrumental'' square visibility amplitude.  Note that $\Vins$ also depends on the spectral profile. It only disappears in the calibration if the calibrator has the same spectral profile as the source.  

In the cases of the Gaussian and top hat transmissions with FWHM $\dwavenum{}$ around central wavelength $\wavenumzero$ and a constant contrast loss $\loss[ij]{}$ in the spectral channel, the smearing is purely real
($\phiband[]{} = 0$) and
\begin{subequations}
\label{eq:easy:vsqPS}
\begin{align}
  \envbandH{}(\alpha)     
  &= \sinc\left(\frac{\alpha}{2\resol{}}\right),
    \label{eq:C:vsqPS}
  \\
  \envbandG{}(\alpha) 
  &= \wideexp{\left(
     -\frac{\alpha^2}{32\resol{}^2\log 2} 
   \right)},
  \label{eq:gauss:vsqPS}
\end{align}
\end{subequations}
where $\resol{} = \wavenumzero / \dwavenum{}$ is the spectral resolution. For small enough baselines, we have shown in Appendix~\ref{ap:smallsmearing} that an exponential formula can be used by properly choosing the value of $\resol{}$.  On real data, $\resol{}$ will need to be set to a value that differs from the spectral resolution in order to account from the departure from Gaussian profile and the wavelength dependence of the contrast.   In practice, a model fit of smeared data may leave it as a free parameter. If high precision is needed, the asymmetry of the spectral band and the slope of $\loss[ij]{}$ give a non zero $\gamma$. Cubic developments for the smearing terms $\textenvband[]{}$ and $\textphiband[]{}$ are given in Appendix~\ref{ap:smallsmearing}.

  %%%% CLOSURE PHASE %%%%

\subsection{Closure phase}
\label{sec:ana:clo}
A triple correlation or its Fourier transform, the bispectrum, or an equivalent method, is generally used to determine the closure phase \citep{LOH83,ROD86}. The determination of the closure phase in direct space uses the phase of the bispectrum, given by:
\begin{equation}
\bispDS[ijk]{} = \intinf 
                    \phasor[ij]{}(\opd[ij](t)) 
                    \phasor[jk]{}(\opd[jk](t)) 
                    \phasor[ki]{}(\opd[\rev{ki}](t)) 
                    \idiff t,
\label{eq:bispDS:1}
\end{equation}
where $t$ is time in the case of linear OPD variations. By substitution of Eq.~(\ref{eq:phasorgen}) into Eq.~(\ref{eq:bispDS:1}) and writing $\textopd[ij](t) = \textdopd[ij] t$
\begin{equation}
  \bispDS[ijk]{} =
                  \sum_{o,p,q} 
                    \intinf
                    \phasor[ij]{o}(\dopd[ij] t) 
                    \phasor[jk]{p}(\dopd[jk] t) 
                    \phasor[ki]{q}(\dopd[ki] t)
                    \idiff t.
\label{eq:def:bispDS}
\end{equation}
It follows from Eqs.~\newrev{(\ref{eq:def:phasor} \& \ref{eq:def:bispDS})} and closure relation $\textdopd[ij] + \textdopd[jk] + \textdopd[ki] = 0$ that
\begin{equation}
\begin{split}
  \bispDS[ijk]{} &\propto
  \sum_{o, p, q}
    \Bigg[
      \exp{i(\phiobj[ij]{o} + \phiobj[jk]{p} + \phiobj[ki]{q})}\\
      &\times
      \intinf 
        \IFTsflux[ij]{o}(\xobj[ij]{o} + \dopd[ij] t)
        \IFTsflux[jk]{p}(\xobj[jk]{p} + \dopd[jk] t)
        \IFTsflux[ki]{q}(\xobj[ki]{q} + \dopd[ki] t)
        \idiff t
    \Bigg].
\end{split}
\label{eq:int:bispDS}
\end{equation}
Using the change of variables $t \rightarrow u = \xobj[ij]{o}/\textopd[ij] + t$, a triple cross-correlation of Fourier transforms can be recognised and expressed as the two-dimensional Fourier transform 
\begin{equation}
  \IFTtd{\ f \ }(\opdvar_1, \opdvar_2) 
    =  \iintinf f(\xi_1, \xi_2) 
            \exp{2\j\pi(\xi_1\opdvar_1 + \xi_2\opdvar_2)} 
             \idiff\xi_1\idiff\xi_2
\end{equation}
of the triple product
\begin{equation}
  \striple[ijk]{opq}(\xi_1, \xi_2) = 
     \sflux[ij]{o}(\xi_1) \sflux[jk]{p}(\xi_2) \sflux[ki]{q} 
     \Big(
     - \frac{\dopd[ij]}{\dopd[ki]} \xi_1 
     - \frac{\dopd[jk]}{\dopd[ki]} \xi_2
     \Big).
  \label{eq:def:striple}
\end{equation}
The bispectrum therefore reads
\begin{equation}
\begin{split}
  \bispDS[ijk]{} \propto 
  \sum_{o, p, q} \Bigg[
    \IFTstriple[ijk]{opq} \Big(
    \phiobj[ij]{o} - \frac{\dopd[ij]}{\dopd[ki]} \phiobj[ki]{q},& 
        \frac{\dopd[jk]}{\dopd[ki]} \phiobj[ki]{q}
        - \phiobj[jk]{p}
      \Big)\\
      &\times \exp{\j\left(\phiobj[ij]{o} + \phiobj[jk]{p} + \phiobj[ki]{q}\right)}
    \Bigg].
\end{split}
\end{equation}
The bandwidth smearing is contained in $\IFTstriple[ijk]{opq}$.  In order to make it clearer we need to introduce several terms.  The triple flux product equivalent (corresponding to $\flux{o}\flux{p}\flux{q}$ in the monochromatic case) is given by 
\begin{equation}
  \triple[ijk]{opq}      = \left| \IFTstriple[ijk]{opq}(0, 0) \right|,
  \label{eq:gen:triple}
\end{equation}
the ``instrumental'' closure phase by
\begin{equation}
  \insphi[ijk]{opq} = \arg \IFTstriple[ijk]{opq}(0, 0), 
  \label{eq:gen:insphi}
\end{equation}
and the smearing by
\begin{equation}
  \smearing[ijk]{opq}(\phivar_1, \phivar_2) = 
      \IFTstriple[ijk]{opq}( \phivar_1 / 2\pi\wavenumzero, 
                            -\phivar_2 / 2\pi\wavenumzero)  
      \,/\,\IFTstriple[ijk]{opq}(0, 0).
  \label{eq:gen:smearing}
\end{equation}
The  ``instrumental'' closure phase is a flux-weighted mean over the spectral channel and thus also depends on the spectrum of the source.  The triple flux product equivalent can be simplified to the triple flux product ($\texttriple[ijq]{opq} \propto \textflux{o}\textflux{p}\textflux{q}$) when the sources have the same spectrum, i.e. $\textsfluxstar{o}(\xi) \propto \textsfluxstar{p}(\xi)$.  Note that the instrumental closure phase cancels out in the calibration only if the sources $o$, $p$, $q$ and the calibrator all share the same spectrum.

With these notations, the bispectrum reads
\begin{equation}
\begin{split}
  \bispDS[ijk]{} \propto \sum_{o, p, q} 
    \Bigg[
      \smearing[ijk]{opq}
    \Big(
    \phiobj[ij]{o} - &\frac{\dopd[ij]}{\dopd[ki]} \phiobj[ki]{q}, 
        \frac{\dopd[jk]}{\dopd[ki]} \phiobj[ki]{q}
        - \phiobj[jk]{p}
      \Big)\\
      &\times\triple[ijk]{opq} \exp{i\left( \phiobj[ij]{o} + \phiobj[jk]{p} + \phiobj[ki]{q}
      + \insphi[ijk]{opq}
                 \right)}
    \Bigg].
\end{split}
\label{eq:gen:bispDS}
\end{equation}
If $\textsmearing[ijk]{opq} = 1$ (no smearing) and $\insphi[ijk]{opq} = 0$ (no bandwidth-related differential phase), the formula is equivalent to the monochromatic case (Eq.~\ref{eq:Bmono} for a binary). In practice,  Eqs.~(\ref{eq:def:striple}, \ref{eq:gen:triple}, \ref{eq:gen:insphi}, \ref{eq:gen:smearing}, \& \ref{eq:gen:bispDS}) allow us to to model fit multiple stellar systems to smeared interferometric data. The knowledge of $\textsflux[ij]{o}$ needed in Eq.~(\ref{eq:def:striple}) can be inferred from calibration fringes obtained on an internal lamp (or a calibrator) as discussed in Sect.~\ref{sec:interferogram}.  

\rev{This modelling} can be further simplified using an analytic description of the bandpass.  In that case, Eqs.~(\ref{eq:gen:bispDS}~\& \ref{eq:bispDS}) can be used for the model fit of closure phases. In our cases of top-hat and Gaussian transmission of FWHM \dwavenum{}, with a linear instrumental phase, we reorder baselines so that $\textdopd[ki]$ has the largest absolute value, and we can assume it negative without losing generality.  Then, the smearing can be simplified to
\begin{subequations}
\label{eq:bispDS}
\begin{align}
  \smearingH[ijk]{}(\phivar_1, \phivar_2) &\propto 
    \sinc\left(
  %    \frac{\phivar_1'}{2\resol{}}
    \frac{\phivar_1 + \phishift[ijk]{}}{2\resol{}}
    \right)
    \sinc\left(
      \frac{\phivar_2 + \phishift[ijk]{}}{2\resol{}}
    \right)
\label{eq:gate:bispDS}
\\
  \smearingG[ijk]{}(\phivar_1, \phivar_2) &\propto
  \exp{ 
    - \frac{
     %   {\phivar_1'} ^ 2 
         (\phishift[ijk]{} + \phivar_1)^2
     % + {\phivar_2'}^2 + 
       + (\phishift[ijk]{} + \phivar_2)^2 + 
          \left(  
            \phishift[ijk]{}
            - \frac{\dopd[jk]\phivar_1 
                  + \dopd[ij]\phivar_2} 
                   {\dopd[ki]} 
          \right)^2 
      }{
        16\resol{}^2\log 2 
        \left(1
        + \big(\frac{\dopd[ij]}{\dopd[ki]}\big)^2 
        + \big(\frac{\dopd[jk]}{\dopd[ki]}\big)^2\right) 
      }
  }.
\label{eq:gauss:bispDS}
\end{align}
\end{subequations}
In the equations above, the ``group delay closure'' is expressed as
\begin{equation}
  \phishift[ijk]{} = 
       \frac{\dopd[ki]\phishift[ij]{} - \dopd[ij] \phishift[ki]{}}
            {\dopd[ki]}
      .
\label{eq:bisp:gd}
\end{equation}
The group delay closure is the consequence of the incorrect centering of the three fringe packets on the three baselines of the telescope triplets.  Because of this de-centering, the centres of these packets are not scanned at the same time.  In order to yield a usable closure phase, there should still be an overlap in the time intervals when the high contrast part of the packets are scanned.  It means that the individual group delays \textphishift[ij]{}, \textphishift[jk]{}, and \textphishift[ki]{}, and thus the group delay closure, should be of the order of a few times the spectral resolution or less ($\textphishift[ijk]{} \lesssim 2\pi\resol{}$).  Since this overlap in time depends on the relative scanning speeds along the baselines, the group delay closure depends on $\dopd[ij]$, $\dopd[jk]$, and $\dopd[ki]$.  

In our analytic approach to the spectral transmission, the instrumental closure phase reduces to a constant term, independent of \newrev{the} source\newrev{s}\begin{equation}
  \insphi[ijk]{} = \insphi[ij]{}(0) + \insphi[jk]{}(0) + \insphi[ki]{}(0).
\end{equation}

Appendix~\ref{ap:disp} explains how to use the Gaussian formula if the the quadratic chromatic dispersion term $\textdisp[ij]{}$ is non zero.

%% BINARY %%%%%%%%%%%%%%%%%%%%%%%%%%%%%%%%%%%%%%%%%%%%%%%%%%%%%%%%%%%%%%%%%%%%

\section{Consequence on companion search}
\label{sec:bin}

\subsection{Bias on the interferometric observables}
\label{sec:bias:observables}

\begin{table}
  \caption{Test case used in \rev{Figs.~\ref{fig:ideal}~\& \ref{fig:phi:jitt}}. For the square visibility amplitude, the first baseline is used.  The spectral resolution is, by definition, the major source of smearing.  In addition, the visibility is slightly impacted by the spectral dispersion $\disp[ij]{}$.  The closure phase is strongly impacted by the group delay closure $\phishift[123]{}$ (indirectly by group delays and OPD modulation speeds) and moderately by the dispersion $\disp[ij]{}$.}
\label{tab:testcase}
\begin{tabular}{ll}
\hline\hline
Binary flux ratio                & 0.6\\
Effective bandpass               & Gaussian\\
Spectral resolution              & lines: 7, 18, 42, contours: 3--100\\
Projected telescope positions    & $(0, B, 0.4B)$\\
\textit{Corresponding baselines} & $(B, -0.6B, -0.4B)$\\
OPD modulation along baselines   & $\dopd[ij] = (\dopd[12], -2\dopd[12], \dopd[12])$\\
OPD bounds                       & $(\pm 25\lambda, \mp 50\lambda, \pm 25\lambda)$\\
Group delays                     & $\phishift[ij]{} = (5, 0, -5)\times2\pi$\\
\textit{Corresponding group delay closure} 
                                 & $\phishift[123]{} = 10\times 2\pi$\\
Spectral dispersion              & $\disp[ij]{} = 0$\\
\hline
\end{tabular}
\end{table}
\begin{figure*}[p]
\subfigure[Square visibility amplitude]{\includegraphics[width=\linewidth]{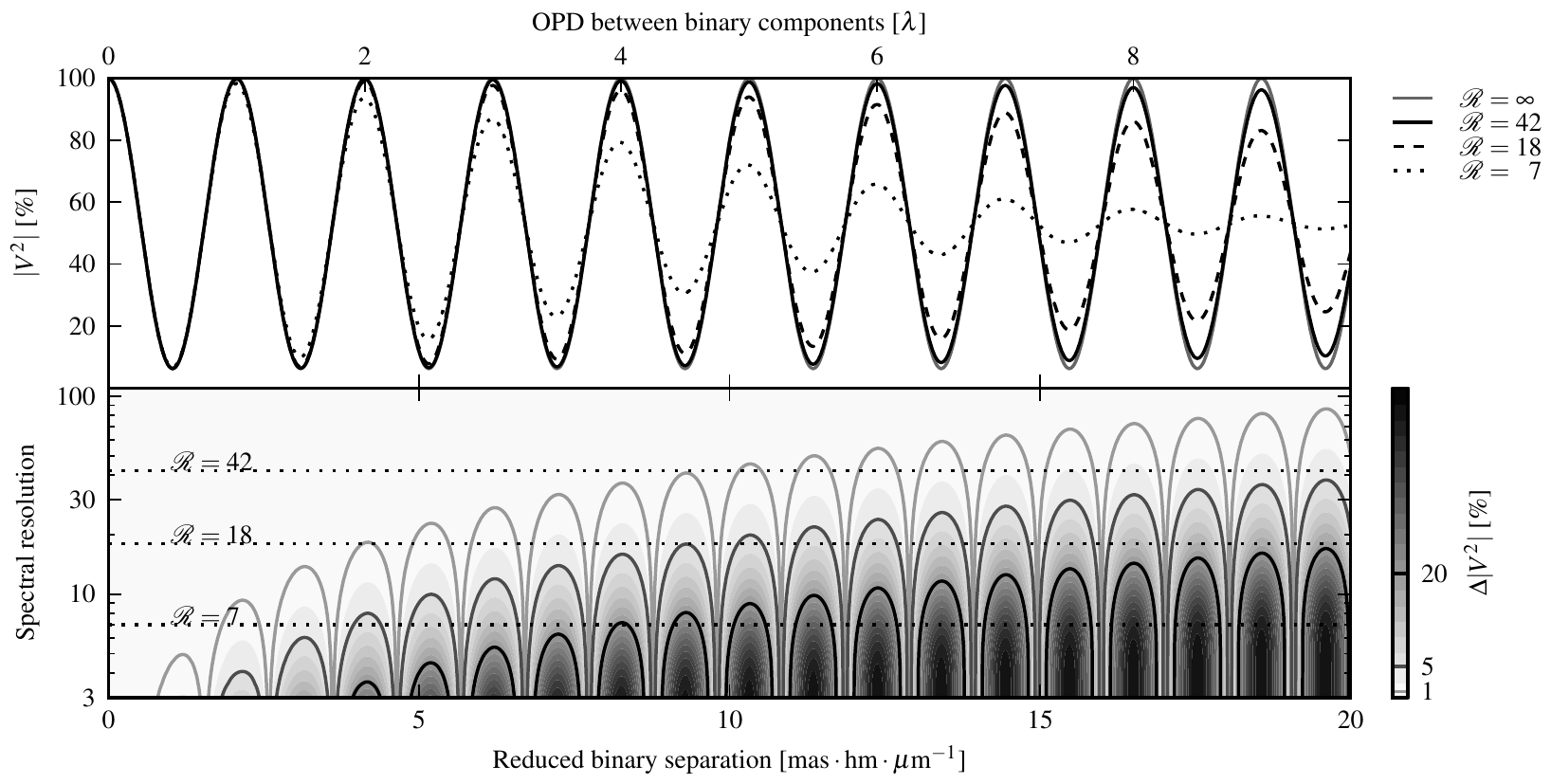}\label{fig:ideal:Vsq}}
\subfigure[Closure phase]{\includegraphics[width=\linewidth]{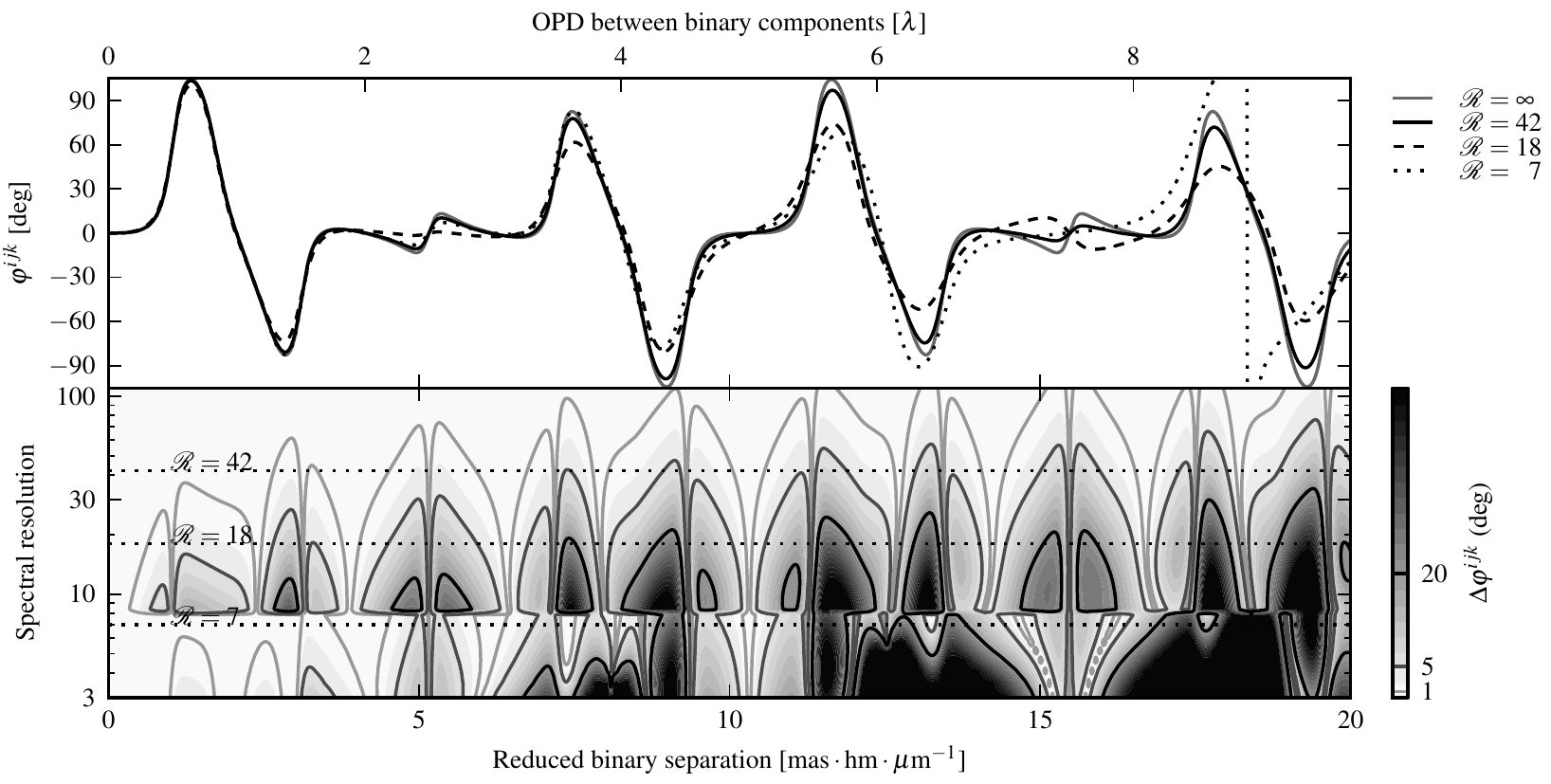}\label{fig:ideal:phi3}}
\caption{Square visibility amplitude (top) and closure phase (bottom) of a binary with flux ratio 0.6 (test case of Table~\ref{tab:testcase}) observed with an interferometer with Gaussian bandpass under ideal atmospheric conditions and baselines $B$, $-0.6B$, $-0.4B$.  In both figures, \emph{top panel:} interferometric observable as a function of binary separation (milliarcseconds at one micron wavelength for a 100\,m baseline) for an infinite resolution and three spectral resolutions approximately matching those of PIONIER.  \emph{bottom panel:} deviation of the smeared observable with respect to the infinite spectral resolution case, shown as contours in the separation-spectral resolution plane.  In the lowest panel, the behaviour change around spectral resolution $\resol{} = 8$ is explained by the transition from the single spectral channel mode (group-delay free in ideal fringe tracking conditions, \rev{since a single fringe packet can be centred around zero OPD, see Appendix~\ref{ap:gd}}) to the multiple channel observation (where \rev{the fringe packets of the different spectral channels are shifted with respect to each other and therefore cannot be simultaneously positioned at zero OPD by the fringe-tracker, see Appendix~\ref{ap:gd}}).}
\label{fig:ideal}
\end{figure*}

The first impact of the smearing is a tapering of the peak-to-peak amplitude of the oscillation of the visibility with baseline, hour angle, or spectral
channel, due to the smearing amplitude $\envband{}$.  The second \newrev{impact} only concerns the closure phase in multi-channel observations\rev{. I}t originates from the imperfect alignment of the fringe packets on baseline triplets, 
as measured by $\phishift[ijk]{}$.  In order to make these influences clearer,
we give in Fig.~\ref{fig:ideal} the interferometric observables of a binary with a high flux ratio 0.6, whose characteristics are given in Table~\ref{tab:testcase}.

\paragraph{Square visibility amplitude.}
Figure~\ref{fig:ideal:Vsq}, top panel, shows the theoretical smearing of the visibility amplitude of a binary as a function of reduced separation $\theta B/\lambda$ (in $\mathrm{mas}\cdot\mathrm{hm}\cdot\mu\text{m}\smash{^{-1}}$) for three different spectral resolutions ($\approx 7, 18, 42$) corresponding to the observing modes available on PIONIER at the VLTI. The lower panel of the figure displays the error on the square visibility occurring from not taking smearing into account, as a function of separation and spectral resolution.   The result is easily generalised to binaries of different flux ratios, as the relative error on the visibility $\Delta|V^2| / |V^2|$ remains unchanged.

\paragraph{Closure phase.}
Figure~\ref{fig:ideal:phi3}, top panel, shows the theoretical closure phase of a binary for three different spectral resolutions ($\approx 7, 18, 42$) corresponding to the observing modes available on PIONIER at the VLTI.  It can be seen at small separations  (5--10\,$\mathrm{mas}\cdot\mathrm{hm}\cdot\mu\text{m}\smash{^{-1}}$) that the intermediate spectral resolution ($\approx 18$) shows more smearing than expected for these separations, in particular more than the broad-band $\approx 7$ observing mode.  The reason lies in \rev{the dispersive elements in the light beams of the interferometer and instrument that decentre fringe packets more in some spectral channels than in others, thus making it impossible to centre all fringes packets at the same time. (see the imperfect centering of some spectral channels of PIONIER in Fig.~\ref{fig:PT} and a description of the group-delay tracking in Appendix~\ref{ap:gd})}. This effect is not seen in the broad band, where \rev{the single fringe packet of each baseline can be centred with a fringe tracker, thus eliminating the group-delay}.  This low-separation smearing approximately scales linearly with separation, as $f\textphishift[ijk]{}\theta/\resol{}\smash{^2}$, where $f$ is the flux ratio of the binary, $\theta$ the separation, and $\textphishift[ijk]{}$ the group-delay closure (This can be obtained analytically by linearising Eq.~\ref{eq:gauss:bispDS} and normalising by the bispectrum of a point-source calibrator.)  At larger separations ($\gtrsim 10\mathrm{mas}\cdot\mathrm{hm}\cdot\mu\mathrm{m}\smash{^{-1}}$ in Fig.~\ref{fig:ideal:phi3}), the closure phase is impacted by a combination of the tapering of the oscillation of the visibility (a purely spectral resolution effect, as seen in the visibility in Fig.~\ref{fig:ideal:Vsq}) and the instrumental phase, the impact is relatively complex, and we can only recommend to use Eq.~(\ref{eq:gauss:bispDS}) to model it.  As an illustration, Fig.~\ref{fig:closim} of Appendix~\ref{ap:closim} compares the closure phase of the three spectral channels of PIONIER for a given configuration of the interferometer, and it is quite clear the the behaviour radically changes with channel and telescope triplet.

The lower panels displays the error on the closure phase occurring from not taking smearing into account, as a function of separation and spectral resolution. The figure shows a sharp discontinuity at resolution $\resol{} = 8$ where the transition occurs from a single spectral channel (where the single fringe packet of each baseline is positioned at zero OPD by an ideal fringe-tracker) to spectrally dispersed fringes (with the fringe packets \rev{of each baseline} that do not align well \rev{because they are shifted with respect with each other by the instrumental phase}). Even for moderately resolved sources, percent precision requires a good enough spectral resolution ($\resol{} \gtrsim 40$ or more), adequate modelling of \rev{bandwidth} smearing, or a good fringe-tracking on a single spectral channel at moderate spectral resolutions ($\resol{} \gtrsim 10$).

\subsection{Retrieving binary parameters}

\begin{figure}
\includegraphics[width=\linewidth]{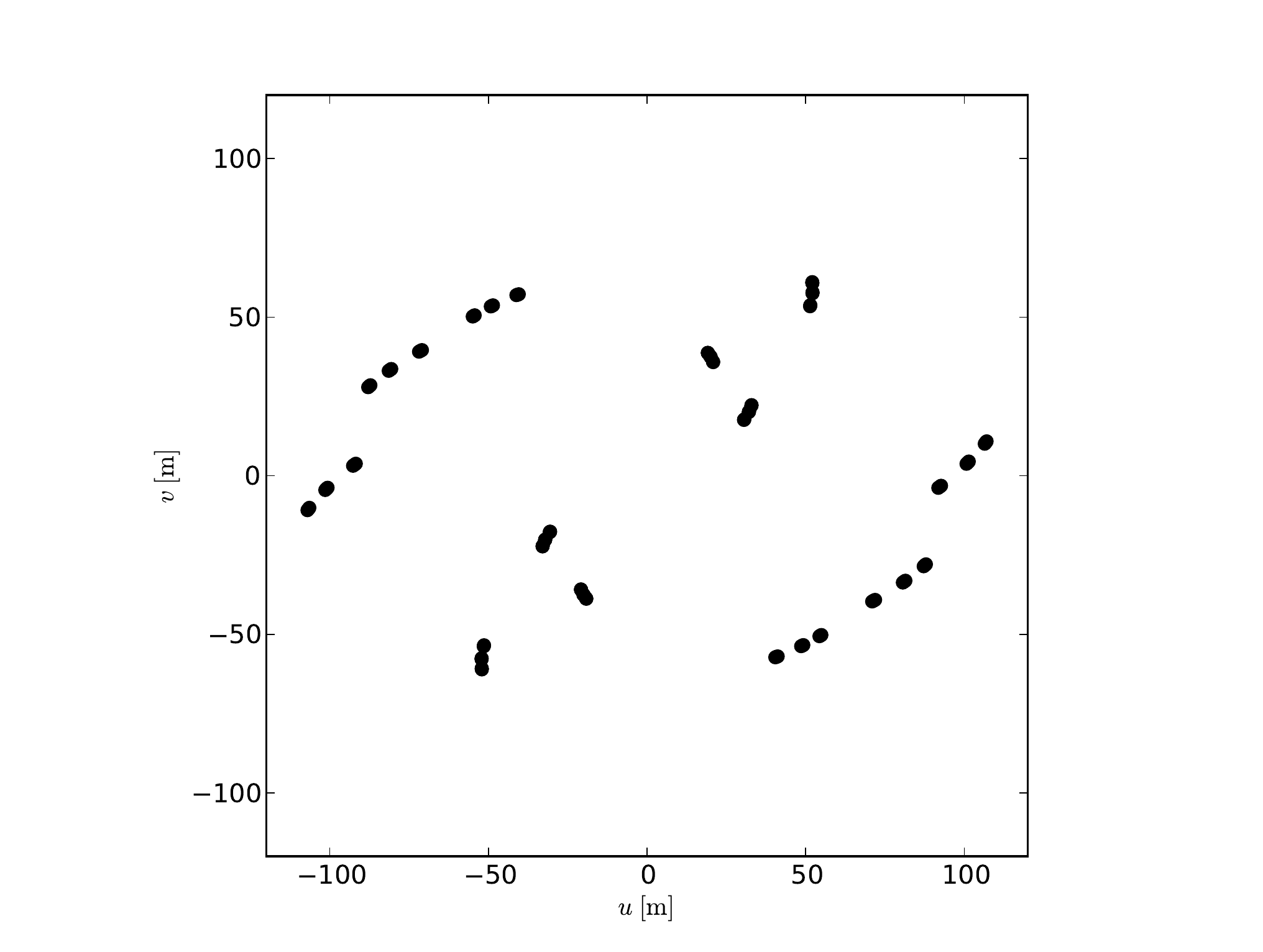}
\caption{$(u, v)$ coverage of a typical 100\,m baseline 4T observation (K0-A1-G1-I1) at the VLTI for an object close to the meridian, with 3 observations over a few hours.}
\label{fig:uv}
\end{figure}

We assess here the bias on the binary parameters that smearing produces. In order to model the data as \rev{realistically} as possible we build synthetic binary fringes corresponding to a \rev{typical scenario}: near-zenith object observed in a sequence of three sets of fringes separated by one hour using a large telescope quadruplet at VLTI (see Fig.~\ref{fig:uv} for $u$, $v$ coverage).  They are obtained from calibration fringes obtained by PIONIER on an internal calibration lamp, which can be considered as a point source observation for our purpose. Then, we feed \rev{these} synthetic data to the PIONIER data reduction software and get visibility amplitudes and closure phases. They are calibrated using simulated fringes of a point-source calibrator.  They are then fit with a binary model to derive the parameters of the binary. In a first step, the model is that of an unsmeared binary (Eqs.~\ref{eq:Vmono}~\& \ref{eq:Bmono}), then we use the smeared model of Sect.~\ref{sec:ana} with Gaussian bandpass (Eq.~\ref{eq:gauss:vsqPS}~\& \ref{eq:gauss:bispDS}). \rev{Additional transmission effects of the VLTI from the telescope up to the internal calibration lamp, positioned after the delay lines, have been ignored: the near-zenith observations we consider here are dominated by PIONIER's instrumental effects (as we discuss in Sect.~\ref{sec:interferogram}).  For non zenithal observations, where the interferometric delay in the delay lines is several tens of metres, the air dispersion in the delay lines becomes a factor of the same order of PIONIER's instrumental phase and can be modelled using Appendix~\ref{ap:gd}.}

\begin{figure*}
\includegraphics{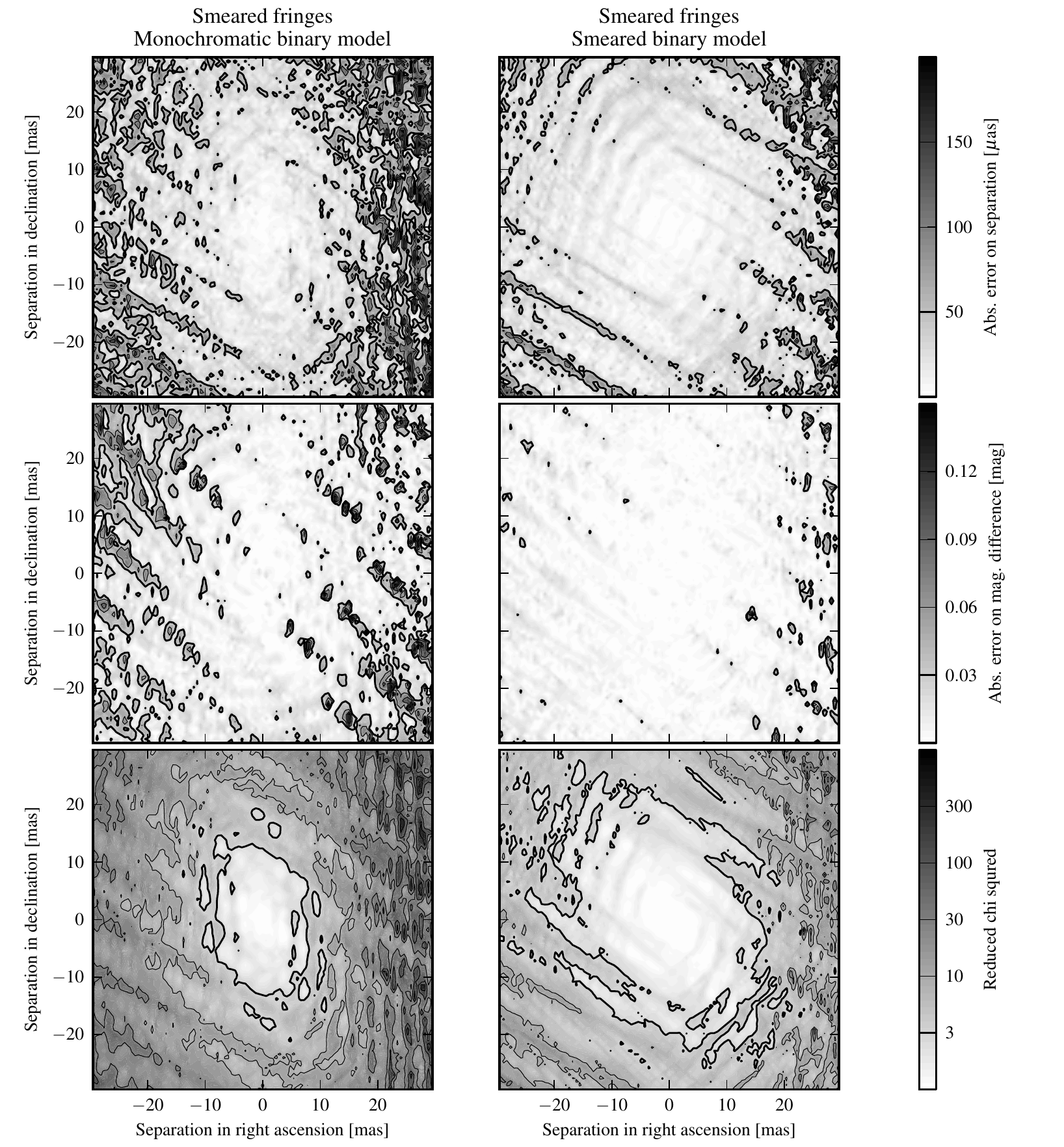}
\caption{Quality of least-squares model fitting of binary parameters to smeared interferometric observables.  These observables are derived from PIONIER synthetic fringes in the 3-channel spectral resolution ($\resol{} \approx 20$) using the data reduction pipeline.  The contour plots give the \newrev{absolute value of the error in the model fits} for each position of the secondary assuming a binary flux ratio of 0.6. \text{Left:} the binary model assumes monochromatic light and absence of smearing. \text{Right:} the binary model assumes a Gaussian bandpass and takes into account the smearing. \text{Top:} \rev{absolute value of the} error on the binary separation. \text{Middle:} \rev{absolute value of the} error on the magnitude difference. \text{Bottom:} reduced chi squares assuming 2\% error on square\newrev{d} visibilities and 20~mrad on closure phases.}
\label{fig:binobs-bias}
\end{figure*}

\begin{figure*}
\includegraphics[width=\linewidth]{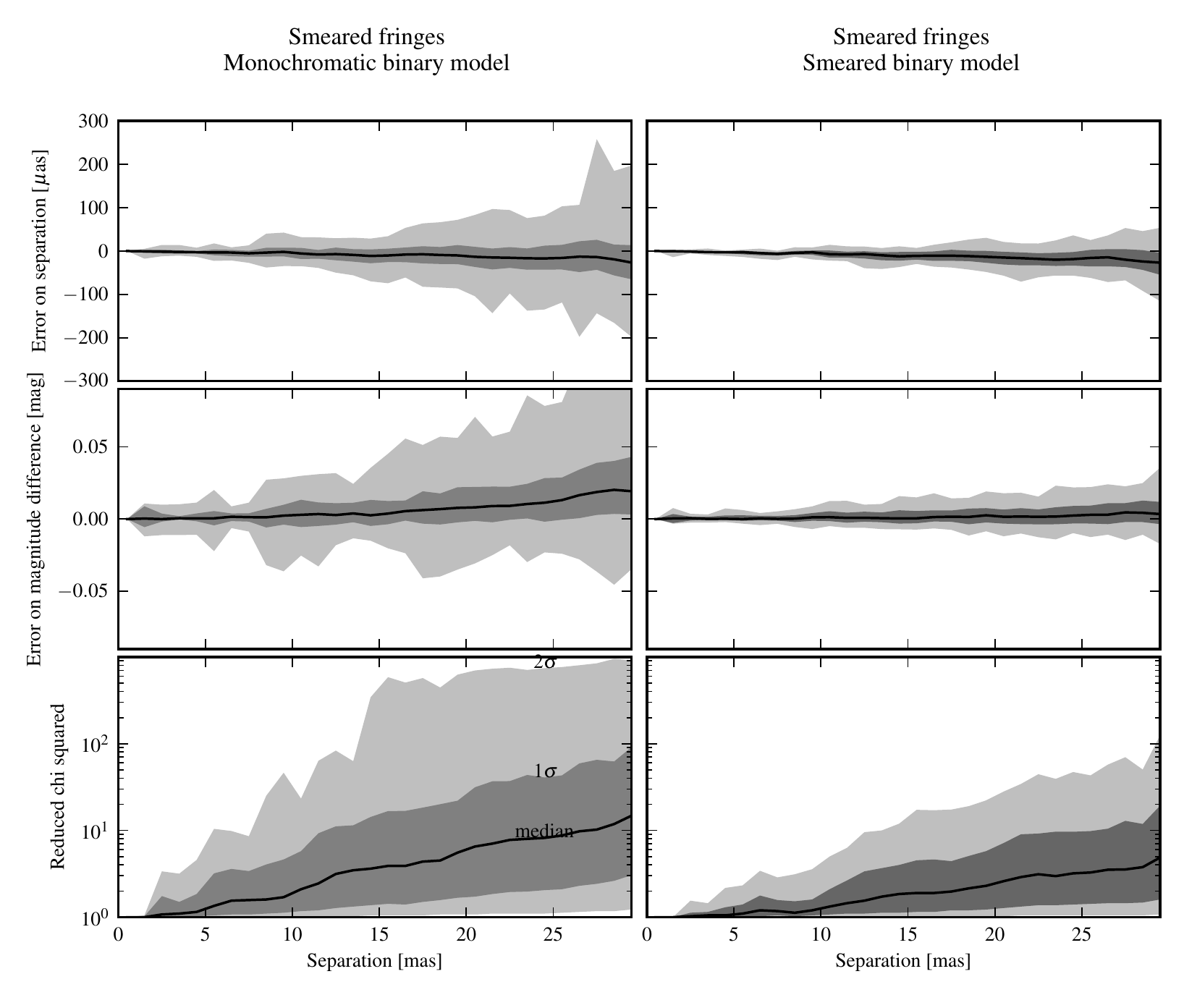}
\caption{\newrev{The solid lines give the median value of the errors on the fitted binary parameters} as a function of binary separation.  \newrev{If non zero and systematically of one sign, the median indicates a bias. The grayed area are the} confidence intervals for the errors (dark gray 1-$\sigma$, light gray 2-$\sigma$). At a given separation, all binary orientations and flux ratios were considered. \text{Left:} the binary model assumes monochromatic light and absence of smearing. \text{Right:} the binary model assumes a Gaussian bandpass and takes into account the smearing. \text{Top:} \newrev{error} on the binary separation. \text{Middle:} \newrev{error} on the magnitude difference. \text{Bottom:} reduced chi square.}

\label{fig:binobs-bias-2}
\end{figure*}

In our analysis, the separations in right ascension and declination are varied from $-30$ to 30\,mas or approximately 10 times the angular resolution the interferometer and the magnitude differences from 0.1 to 3.3 (flux ratios from 0.05 and 0.95).  For each point triplet of parameters, the difference between the fitted values and the input gives us the bias on the binary position and magnitude difference.  The reduced chi square was determined assuming a 2\% accuracy on visibilities and 20\,mrad on closure phases typical of single-mode instrument performances on bright objects (like PIONIER).  Figure~\ref{fig:binobs-bias} shows the \rev{absolute values of the errors} and reduced chi square at each separation and position angle at the given magnitude difference of 0.55 (flux ratio of 0.6). In Figure~\ref{fig:binobs-bias-2}, we \rev{consider possible biases and give} the median value of the \rev{error with} its confidence intervals for a given binary separation, considering all the position angles and flux ratios at that separation.

\paragraph{Smearing-free binary model.}  A binary model with the classical expression for the visibility amplitude and closure phase (Eqs.~\ref{eq:Vmono}~\& \ref{eq:Bmono}) is fitted to synthetic PIONIER data with the three-channel spectral resolution.  
The left panel of Fig.~\ref{fig:binobs-bias} displays from top to bottom the \newrev{absolute value of the error} on the secondary's position, the \newrev{absolute value of the error} on the magnitude difference, and the reduced chi square for errors of 2\% and 20\,mrad on individual measurements of square visibility amplitudes and closure phases respectively.   We checked that the results for other flux ratios are similar. The \newrev{errors (with median value and confidence intervals)} for the parameters are given in Fig.~\ref{fig:binobs-bias-2} (left panel) as a function of separation when the flux ratio is allowed to vary between detectable limits (0.05 to 0.95). \newrev{The median value of the error indicates a bias, if it is non zero and consistently of one sign.} 

The main impact of the smearing is a degradation of the goodness of fit at all separations, followed by errors on the flux ratio and separation at moderate separations, and a clear bias of both observables at larger separations. In our models, the secondary is dimmer than the input of the simulation \newrev{more often than not} and the separation tends to be smaller \newrev{more often than not}. \newrev{(For instance, the confidence intervals on the errors of Fig.~\ref{fig:binobs-bias-2} show that the error on the separation is approximately 5 times more likely to be negative than positive at a separation of 30\,mas.)} The \newrev{apparent dimming of the secondary} is easily explained by the tapering of the fringe contrast that occurs due to smearing. The \newrev{bias on separation} is independent of smearing as we will see later on. 

Even at moderate separations (5--10\,mas) the reduced chi square is around 3.  However, the errors on the flux ratio and positions become significant (50\,$\mu$as and 20\,mmag) only at higher separations ($\gtrsim 15$ mas), as Fig.~\ref{fig:binobs-bias}.  At first sight, it seems to contradict the trend of Sect.~\ref{sec:bias:observables}. In that section, we have found a significant smearing of the closure phase at small separations, as a result of the imperfect centering of fringe packets in an observation with multiple spectral channels.  We easily reconcile these findings by noting that, as an average over the spectral band, the group delay is zero, i.e. both ends of the bands have group delays of same magnitude but opposite signs; thus their respective impacts on the observables approximately cancel out in the fit. The deviation of the individual spectral channels from the average over the band still explains the larger chi square. (Fig.~\ref{fig:closim} in Appendix~\ref{ap:closim} shows how the closure phases are impacted differently for the three spectral channels of PIONIER in low resolution mode.) 

\paragraph{Smeared binary model.}  We performed similar fits to synthetic smeared fringes of a binary \rev{by} using the Gaussian formulas for the smearing (see Sect.~\ref{sec:ana}).  The \newrev{absolute values of the errors} on the position and flux ratio are given for a binary with a flux ratio of 0.6 in the right panel of Fig~\ref{fig:binobs-bias}. The \rev{errors} on the position and magnitude difference, and the quality of the fit are given in the right panel Fig.~\ref{fig:binobs-bias-2} for a wide range of flux ratios.  \newrev{In Fig.~\ref{fig:binobs-bias-2}, the median value of the error indicates a bias if it is non zero and consistently of one sign.}

Taking the smearing into account eliminates most of the errors and bias on the flux ratio. It also largely improves the quality of the fit, with a reduced chi square of 3 found at significant separations ($\gtrsim 15$\,mas) in \rev{most cases}.  The errors on the separation are improved at all separations but \rev{the bias remains at larger separations}. We \rev{have found that the bias is related} to the uncertainty on the effective wavelength of the interferometer, which varies by $\approx 0.1$\% across baselines on PIONIER; this phenomenon is independent \rev{of} our adequate modelling of the smearing. \rev{It is difficult to calibrate in the first place, because a deviation of the pie\newrev{z}o scan speed from its nominal value has exactly the same observable consequence. (We note that including a proper spectral calibration in the instrument would solve for this problem.)}  At 30 mas of separation, \rev{a 0.1\% bias translates into 30$\,\mu$as}, which is what we indeed find: \rev{the solid lines in the top panels of Fig.~\ref{fig:binobs-bias-2} show this bias both in the monochromatic model and the smeared one.} At specific binary parameters, seen as high \rev{error} values islands on Fig.~\ref{fig:binobs-bias}, \rev{the discrepancy} originates from the difference between the smeared visibility and the Gaussian model: This happens close to smearing-induced phase jumps (see Fig.~\ref{fig:closim} of Appendix~\ref{ap:closim} for a comparison between Gaussian smearing and simulated values). High contrast binaries \rev{do not feature these phase jumps} and are not impacted.  For precision work \rev{of high to moderate flux ratio binaries, we strongly recommend to discard closure phases} close to predicted jumps.

%%%%%%%%%%%%%%%%%%%%%%%%%%%%% ATMOSPHERE %%%%%%%%%%%%%%%%%%%%%%%%%%%%%%%%%%%%

\section{Modelling the atmosphere}
\label{sec:atm}
\label{sec:atm:temp}

The estimators of the interferometric observables have been chosen to be mostly immune to atmospheric biases in the typical interferometric regime of a moderately resolved source, \rev{i.e. when bandwidth smearing can be ignored}. In this section, we investigate possible biases when \rev{bandwidth smearing becomes significant}, as \citet{ZHA07} did for IOTA's closure phase estimator.

For temporal scanning, it is possible to write the differential piston---the variable differential phase induced by the atmosphere---as a function of OPD since time and OPD are linked \citep[see for instance][]{jitter}.  The jittered coherent flux can be expressed as a function of the ideal coherent flux
\begin{equation}
  \phasorjitt[ij]{}(\opdvar) = 
     \phasor[ij]{}(\opdvar + \piston[ij](\opdvar))
     \wideexp{\left[
       -\frac16 \left(
         \pi\wavenumzero 
          \pderiv{\piston[ij]}{\opdvar}(\opdvar)
        \right)^2\right],
     }
  \label{eq:coherjitt}
\end{equation}
\rev{where $\textpiston[ij]$ is the atmospheric differential piston on baseline  $ij$.} The exponential term is the contrast loss due to piston variation during the integration, of the order of one millisecond for one OPD step of a temporal scan.  It bears the assumption that the spectral envelope of the fringes does not have features as sharp as the fringe frequency and that the integration during one OPD step is fast enough (of the order of \rev{a} millisecond in practice) to allow for a linearisation of piston.  

\subsection{Orders of magnitude}
\label{sec:atm:om}
An analytic approach to the atmospheric turbulence can be taken, using the 
assumption that scanning is fast enough for the piston to vary linearly during
a sub-second scan, i.e. $\textpist[ij]{} = \textpist[ij]{0} + 
\textpist[ij]{1} \textopd[ij]$, where $\textpist[ij]{0}$
is the group-delay tracking error and $\textpist[ij]{1}$ a rate of piston 
variation during scan. $\textpist[]{0}$ and $\textpist[]{1}$ are random variables when statistics over a large number of scans are derived. Using this approach, the coherent flux is:
\begin{align}
\begin{split}
  \phasorjitt[ij]{} (\opd[ij]) &= 
    \sum_o
    2 \IFTsflux{o}(\xobj[ij]{o} + (1 + \pist[ij]{1})\opd[ij] + \pist[ij]{0})       
    \\&\qquad \times \exp{
        i\phiobj[ij]{o} 
     + 2i\pi\wavenumzero[(1+\pist[ij]{1})\opd[ij] + \pist[ij]{0}]
     - \frac16 (\pi\wavenumzero\pist[ij]{1})^2 
    }.
\end{split}
\label{eq:atm:phasor}
\end{align}
This approach can be used to determine the orders of magnitude of the atmospheric effects.

\paragraph{Visibility.} 
The piston variation term $1 + \textpist[ij]{1}$ comes as a product of the OPD variable in Eq.~(\ref{eq:atm:phasor}), so we recognise it as a scaling factor.  $\textpist[ij]{0}$ is a mere shift of the central OPD and has no impact---the square visibility does not depend on centering.   Therefore, we can link the jittered visibility to the ideal case: 
\begin{equation}
  \vsqPS[ij]{\text{jit}} = \frac{1}{1+\pist[ij]{1}} \vsqPS[ij]{\text{ideal}}
  \wideexp{-\frac13 (\pi\wavenumzero\pist[ij]{1})^2}.
\end{equation}
The impact of atmospheric jitter is independent \rev{of} the geometry of the source and, thus, smearing.  For all separations it can be calibrated out if science target and calibrators are observed with similar atmospheric conditions.

\paragraph{Closure phase.}  The group-delay tracking term $\textpist[ij]{0}$ can be seen as a fringe shift that adds to the predicted fringe position $\textphishift[ij]{} \rightarrow \textphishift[ij]{} + 2\pi\wavenumzero\textpist[ij]{0}$ and the linear variation of the piston can be seen as a scanning velocity change $\textdopd[ij] \rightarrow \textdopd[ij](1 + \textpist[ij]{1})$.   With these substitutions, the formulas of Sect.~\ref{sec:ana:clo} can be used directly to determine the jittered closure phase. As we have seen, the predominant impact of the bandwidth smearing on the closure phase is the fringe decentering $\textphishift[ij]{}$, so we expect the group-delay tracking errors to be the main source of bias. 

\subsection{Numerical modelling}

\begin{figure}
\includegraphics[width=\linewidth]{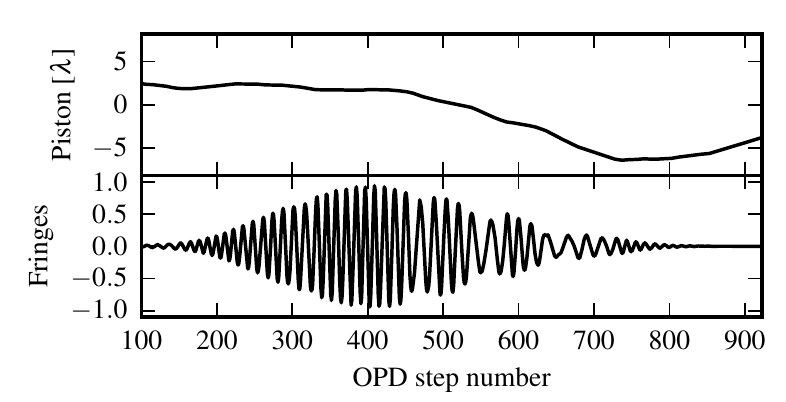}
\caption{One of the simulated temporal scans. The deformation of the envelope is correlated with the piston slope and the accordion-like features to variations of its slope.  \textit{Top:} piston; \textit{Bottom:} simulated fringes.}
\label{fig:interf:jitt}
\end{figure}
In the high frequency regime the pistons at the different stations can be considered as uncorrelated when the baselines are larger than the outer scale of turbulence $\mathcal{L}_0$ \citep{KEL07}.  With a median value $\mathcal{L}_0 = 22$\,m at Paranal \citep{MAR00} baselines of the medium and large telescope quadruplets used with PIONIER normally fulfil the criterium.  At other sites, for smaller baselines, or under relatively uncommon atmospheric conditions at Paranal, the pistons can be correlated. This correlation decreases the amount of atmospheric jitter for given coherence time and seeing, which in turns tends to decrease the bias on the interferometric observables. Therefore, we model the random piston $\piston[i](t)$ using its spectral density
\begin{equation}
  \DFT{\piston[i]}(\nu) = A\nu^{-B} \exp{\j\Phi^i(\nu)},
\end{equation}
where $A$ and $B$ are constants and $\Phi^i(\nu)$ is chosen randomly for each sampled temporal frequency $\nu$.  For Kolmogorov turbulence, the fast scan ($\ll 1$\,s) regime has $B = 17/6$ \citep{CON95} but there is \rev{experimental evidence \citep{DIF03}} that the slope is not as steep at VLTI, \rev{with simulations by \citet{ABS06} explaining it in terms of the piston induced at high frequency by the adaptive optics (imperfect) correction \citep[``bimorph piston'', see][]{VER01} and wavefront errors produced by the injection into single-mode waveguides \citep[``coupled piston'', see][]{RUI01}. \citet{LIN99} have also measured a deviation from the Kolmogorov behaviour at PTI.} We used $B = 2$, which experimentally reproduces well the accordion features of temporal scans obtained under below average atmospheric conditions (see Fig.~\ref{fig:interf:jitt}). We normalise $A$ to match the group-delay tracking rms in the differential piston $\piston[ij] = \piston[j] - \piston[i]$.

By substituting in Eq.~\ref{eq:atm:phasor}, we perform a numerical integration of Eqs.~(\ref{eq:def:vsqPS}~\& \ref{eq:def:bispDS}) and obtain the jittered
visibility amplitude and closure phase.

\begin{figure*}
\includegraphics{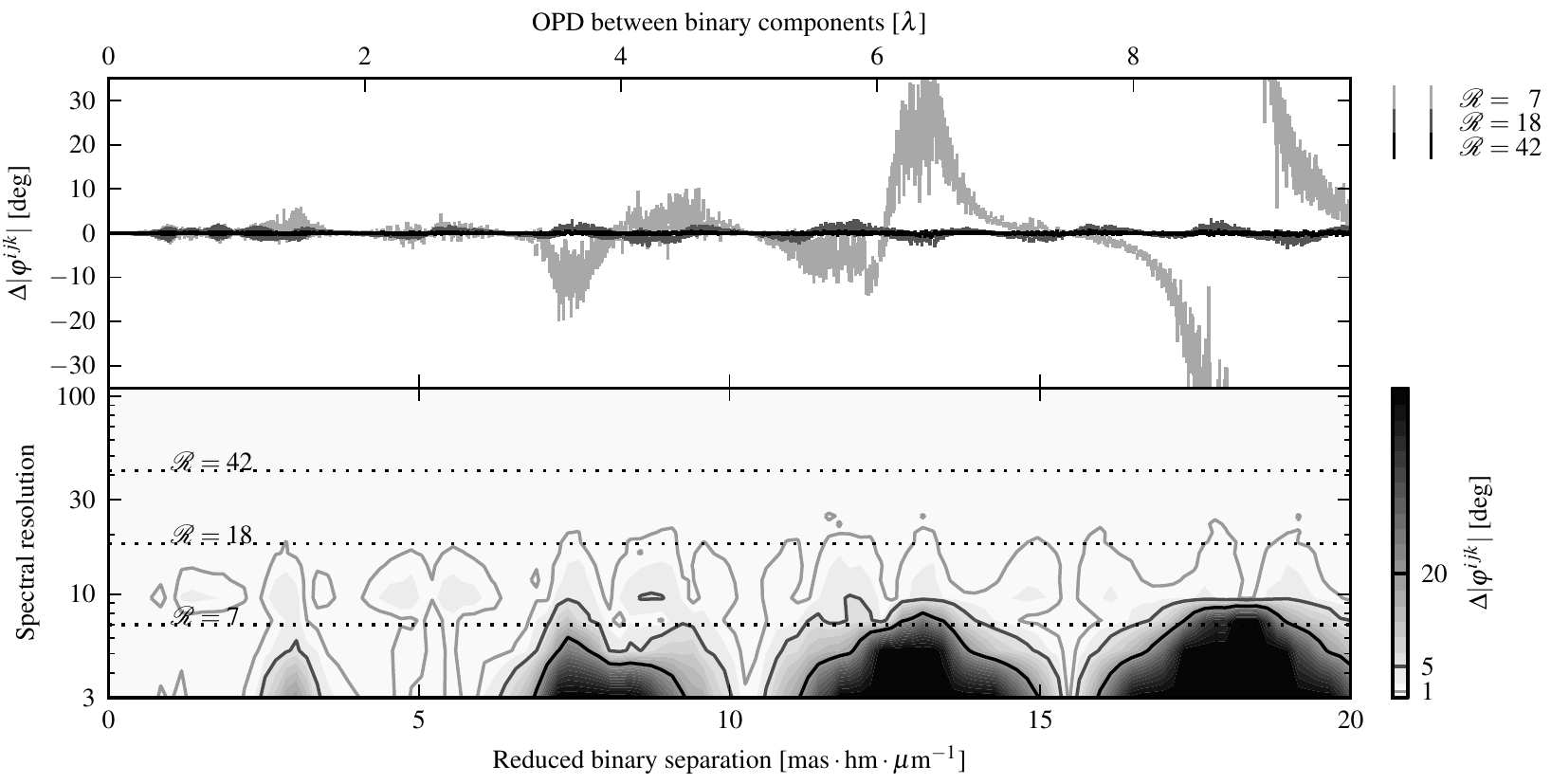}
\caption{Bias on the closure phase resulting from atmospheric piston in temporal scans, assuming that the static smearing is correctly modelled. The x-axis shows the reduced binary separation in milliarcseconds-hectometres of baselines per micron of wavelength (below) or the OPD between binary components (above). \textit{Top:} bias and statistical errors for three spectral resolutions corresponding to PIONIER at the VLTI. \textit{Bottom panel:} bias in the spatial resolution-spectral resolution plane. The bias decreases quickly with spectral resolution.}
\label{fig:phi:jitt}
\end{figure*}

\subsection{Bias on the observables} 
As we have seen in Sect.~\ref{sec:atm:om} there is little bias of the atmosphere on the square visibility amplitude and we could confirm it numerically. However, the bias can be substantial on the closure phase. Figure~\ref{fig:phi:jitt} displays in its top panel the bias on the closure phase of our test-case binary as a function of separation, for the three spectral resolutions $\resol{} = 7$, 18, 42 corresponding to PIONIER's modes.  \rev{For each separation, baseline, and spectral resolution considered in the simulation}, 100 random scans with a \rev{remaining scatter of the fringe tracking of $6\lambda$ (typical value by average conditions) have been generated.  The closure phase on the telescope triplet is the average closure phase of the scans.  To better identify the biases}, the closure phase of a \rev{jitter-free observation} has been subtracted from the results. In the lower panel of the figure, the bias on the phase is given in the separation-spectral resolution plane.  As one can see, the impact of the atmosphere is very important at low resolution but quickly vanishes for $\resol{} \gtrsim 20$. For three spectral channels across a typical IR band, the error on the phase is at most a few degrees or less. 

\section{Discussion \& Conclusion}

\subsection{Impact of the instrument and visibility estimator}

As already discussed by \citet{PER05}, the square visibility amplitude is impacted differently for different estimators that otherwise would be equivalent in the absence of smearing. Not only is the amount of smearing different but the behaviour can be changed.  Because it is a popular recombination method and it illustrates this argument, we have given the formulas for the smeared complex visibility of a time-modulated ABCD recombiner in Appendix~\ref{ap:ABCD}. In Sect.~\ref{sec:ana}, we have seen that the square visibility amplitude is not impacted by the fringe centering in full scans processed by Fourier analysis : in Eq.~(\ref{eq:gen:V2smearing}), smearing is independent \rev{of} absolute source position---only on source distances $\textphiobj[ij]{o} - \textphiobj[ij]{p}$---and group delay $\textphishift[ij]{}$. Conversely, the ABCD visibility estimator shows explicit dependence on $\textphiobj[ij]{o}$ and $\textphishift[ij]{}$ (see for instance Eq.~\ref{eq:ABCD:gauss:smearing}), and this propagates to the square visibility estimator.  

Also, we have clearly put in evidence that instrumental features such as the OPD modulation scheme \rev{(ABCD or Fourier mode, stroke speeds on the different baselines)} or the chromatic dispersion have a strong impact on the closure phase. In particular, the smearing behaviour of the closure phase of PIONIER (Fig.~\ref{fig:closim}) shows different trends on different triplets or different spectral channels: on one hand, different telescope triplets are impacted differently because of the different OPD modulations; on the other hand, different spectral channels of the same triplet behave in different manners, as a consequence of different chromatic signatures.  While the square visibility amplitude did not show a strong dependence on instrumental signature for full scans processed by Fourier analysis (Sect.~\ref{sec:ana}), this is not necessarily the case.  For instance, a time-modulated ABCD method displays impact for both visibility and phase (see Eq.~\ref{eq:ABCD:gauss:smearing} in Appendix~\ref{ap:ABCD}).

\rev{We therefore stress} that each data reduction pipeline and each instrument require their own modelling of the smearing.  In this paper, we have provided a generic formalism which can be used as is for VLTI/PIONIER and probably with little adaptation to other instruments that temporally scan most of the fringe packet.  

\subsection{When only part of the fringe packet is sensed}

Also, our developments make the implicit assumption that most of the flux of \newrev{the} fringe packet is measured, i.e. that the OPD range is significantly larger than the FWHM of the fringe envelope. Actually, our developments still hold if the centres of the fringe packets originating from the different parts of the source are scanned but the extremities of the fringe packet are cropped, providing that the cropping is not too aggressive. \rev{In the case of PIONIER, the partial cropping on some baselines does not prevent a good agreement between simulated fringed and our analytic development, as Fig.~\ref{fig:closim} shows.}   

However, it is clearly not the case in the ABCD method when a fringe-tracker locks the recombiner on the ``central'' fringe \citep[e.g][]{SHA80}.  While the smearing can be derived theoretically for this method (see Appendix~\ref{ap:ABCD}), \rev{its magnitude will depend on the location of the fringe (i.e the OPD) onto which the fringe tracker locks. In the aforementionned Appendix it is shown that the visibility depends on the position of a source which in turns depends on the value of the group delay \textphishift[ij]{} (see Eq.~\ref{eq:ABCD:beta}). For relatively compact objects, the fringe tracker locks onto the brighter fringe or a local zero of the group delay and possible biases are calibrated out when observing an (almost) point-like calibrator under similar conditions.  When a source is smeared, the fringe tracker does not necessarily behave in the same manner on source and calibrator, since there is no longer an obvious location of a central fringe (e.g. in the extreme case of a double fringe packet, it may lock on either packet). Therefore,} it is quite likely that instruments sensing the central fringe of sources more resolved than a few beam widths \rev{(i.e. a few times the resolution power of the interferometer) will lead to altered measurements}, unless \rev{(a)} a high spectral resolution \rev{is used ($\resol{} \gg \textphishift[ij]{}$ in Eq.~\ref{eq:ABCD:beta})} or \rev{(b) the fringe tracking scheme can be modelled with enough detail to know on which part of a given smeared fringe packet it locks}.  In particular, instruments that target high accuracy astrometry with the ABCD method like GRAVITY \citep{GRAVITY} and PRIMA \citep{PRIMA} will require that both the tracking reference and the science target are not very resolved.

\subsection{Image reconstruction}
Our approach clearly targets parametric analysis, by providing formulas to model fit interferometric data by systems of compact sources.  Image reconstruction however, usually relies on the Fourier relation between visibility and image, a relation which is broken in finite bandwidth.  Thus, image reconstruction is made difficult as \cite{BRI89} already noted in radio interferometry.

\subsection{Dealing with bandwidth smearing in practice}
The angle of attack of radio astronomers to limit bandwidth smearing
(see e.g \citet{BRI89}), is to restrict its effects either by
increasing the spectral resolution to optimise the interferometric
field of view or centering the phase tracking delay optimally to
reduce the radial spread. Optical interferometry users do not have
necessarily such a flexibility. One of the important differences
between the wavelength regimes is that, in the optical, because the
arrays have \rev{many fewer} telescopes, most of the users do not actually
reconstruct images but rather model directly the interferometric
observables. This \rev{has} been done to an extreme level of precision
where visibilities are measured to a fraction of percent
\citep[e.g.][]{Absil:2008} and closure phases to a fraction of a degree
\citep[see e.g][]{Zhao:2011}. The particularly large impact of the
smearing, even for moderately resolved sources, undermine the idea
that the parameters for a large number of objects might be derived
effortlessly using the traditional techniques.

It therefore appears reasonable to adopt a two step strategy to deal with
bandwidth smearing first by \emph{limiting the static instrumental smearing
by design} and secondly by \emph{operating the instrument under
conditions that allow a proper modelling of the induced biases}.

\emph{Limiting the instrumental smearing.} We have seen that the ``group delay
closure'' is the major contributor to a static smearing effect in the closure
phase \rev{for instruments that operate in Fourier mode}; it depends on the
group delays and the OPD modulation scheme. The scanning speed scheme can be
chosen so as to minimise the average group delay closures. For the
\rev{temporal ABCD, visibility amplitudes and closures phases are directly
impacted by the group delay, and this mitigation can longer be used.  Since the
group delay is mostly produced by a static chromatic dispersion in the instrument (waveguides, optical elements), an} integrated approach
to differential dispersion and birefringence compensation can be attempted as
discussed in \citep{LAZ12}. Solutions exist that can provide guided or free
space optics instrument with dispersion compensation \citep{Vergnole:2005}.
\rev{Correcting the air dispersion in the delay lines in real time may prove
more difficult to implement than static correction of the dispersion in the optical elements, so that evacuated delay lines are probably part of the solution for larger baseline lengths ($\gg 100$\,m) \newrev{and at shorter wavelengths where the air dispersion is larger}.}

\emph{Modelling the biases.} We have shown that bandwidth smearing can be
modelled provided that, a moderate spectral resolution is used (the first
obvious step) \rev{and} the \rev{estimators of the observables are properly
calculated}.  In very low spectral resolution or in full-band ($\resol{} \sim
5$) observations atmospheric effects must also be decently constrained. For the
latter, initial studies \citep[e.g.][]{LIN99,DIF03} have shown the correlation
between atmospheric turbulence and low frequency statistics of piston but these
are not necessarily well adapted to the sub second exposure
\citep[e.g.][]{ABS06}. Dedicated further characterisation of piston statistics
\rev{vs. monitored atmospheric properties} would be needed. In summary, the
ultimate tool to obtain a smeared source's \rev{properties} will simulate the
instrumental visibility numerically taking the instrumental signatures, in
particular a dedicated spectral calibration, and the atmosphere into account.

\subsection{Concluding remarks}

\beginrevision
Optical interferometry is increasingly used for precise measurements of high flux ratios and/or separation. Application of this precision techniques range from the detection of hot dust components around debris-disc host stars or the search for direct detection of hot Jupiters to the accurate astrometry of binary systems in search of precise mass determination. 

We have focused our work on a rarely studied effect that can alter significantly these astrophysical measurements, the so-called the bandwidth smearing. This bias-inducing phenomenon arises from the wavelength-dependence in the characteristics of the instrument, the atmosphere, and the source. We have modelled its impact by analysing its influence on the instrumental fringe contrast and determined how it alters the visibility amplitudes and closure phases. The magnitude of this effect will depend, for a given instrument, on the spectral resolution and the extension of the observed field of view and in some cases on the atmospheric piston.

We have demonstrated analytically how to calibrate for this degradation in the context of popular temporal fringe scanning instruments and applied this analysis to the specific case of binary systems by computing the error or biases induced on the separation vector and flux ratio.

We have further discussed ``real-life'' constraints such as the influence of the atmospheric piston, the use of different fringe encoding schemes or the imperfections of the fringe tracking quality.  We believe that the current analysis can be used with little effort to correct for potential bandwidth smearing biases in almost any astrophysical case.
\endrevision

\section*{Acknowledgements}
\rev{We would like to thank an anonymous referee and Chris Haniff who helped us to improve this paper.  This research has made use of NASA's Astrophysics Data System, the free softwares maxima, Yorick, and python. It has been supported by Comit\'e Mixto ESO-Chile and Basal-CATA (PFB-06/2007).}

{\footnotesize
\bibliography{lachaume}}
        
\appendix

%%%%%%%%%%%%%%%%%%%%%%%% ADDITIONAL FORMULAE %%%%%%%%%%%%%%%%%%%%%%%%%%%%%%%

\section{Additional formulas}

 %%% COMPACT BINARY %%%

\begin{figure*}
  \includegraphics[width=\linewidth]{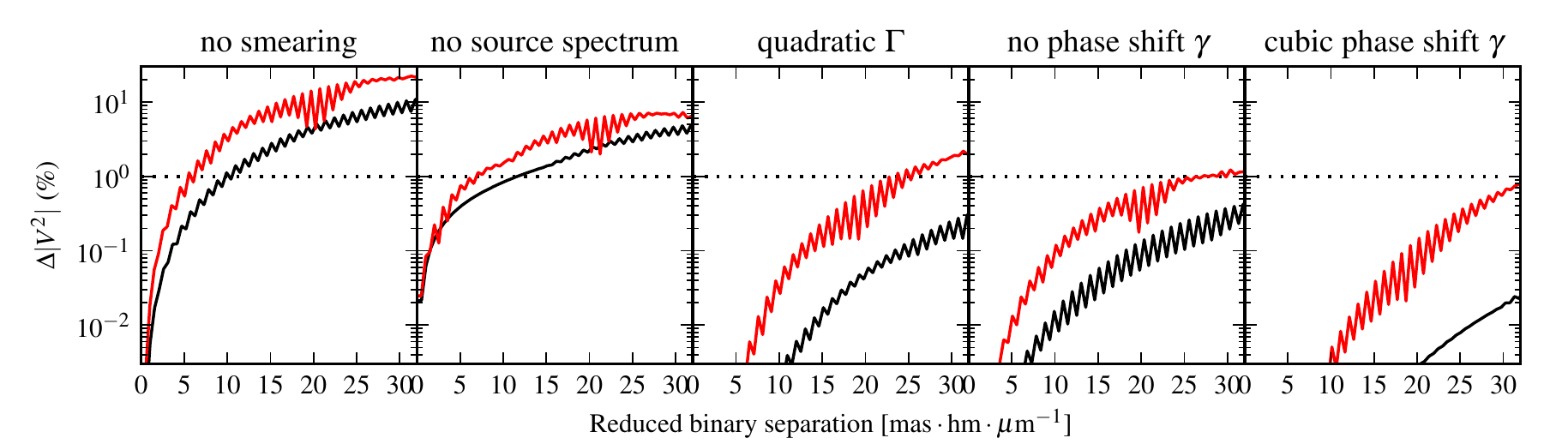}
  \caption{Error budget of the smearing for the two spectral
    configurations of PIONIER in the H band (upper curve: 7 channels;
    lower curve: 3 channels).  The x-axis gives
    the binary separation in units of milliarcsecond for a 100\,m 
    baseline per micron of wavelength. Each figure displays the maximum error on the square visibility amplitude resulting from the following approximations: not
    taking into account the smearing itself (up to 20\% has been
    considered, more is almost the double packet),  the source's
    spectrum (up to 5\% difference for spectral indices 0 and ${-}1$), the non quadraticity of the smearing amplitude $\Gamma$ (up to 2\%), the phase shift $\gamma$  (up to 1\%) and the departure of the phase shift from the
    cubic approximation (less than 1\%).  For most applications, where
    the visibility has a stability of a few percen\rev{t}, a quadratic or
    exponential smearing is accurate enough.}
    \label{figap:accuracy}
\end{figure*}

 %%% RESOLVED SOURCES %%%

\subsection{System of compact, resolved sources}
\label{ap:syscomp}
We consider sources indexed by $o$ with with complex visibilities (amplitude and phase) given for \rev{baseline} $ij$ by
\begin{equation}
  \svis[ij]{o}(\sigma) = \visamp[ij]{o}(\sigma) 
                         \wideexp i\visphi[ij]{o}(\sigma),
\end{equation}
when taken with respect to their nominal centre. The interferograms of the compact sources are not individually impacted by bandwidth smearing, so it is straightforward to see that the developments are similar to those of Sect.~\ref{sec:ana}, with $\textsflux{o}$ substituted to by $\textsflux{o}\textsvis[ij]{o}$ in the complex coherent flux (Eq.~\ref{eq:def:phasor}).   Under the assumption that the individual visibilities are approximately constant over the bandwidth, formulas for the visibility, differential phase, and closure phase can be derived from Eqs.~(\ref{eq:gen:vsqPS} \&~\ref{eq:gen:bispDS}) using the following substitutions: 
\begin{align*}
  \phiobj[ij]{o}    &\rightarrow \phiobj[ij]{o} + \visphi[ij]{o},\\
  \norm[ij]{op}     &\rightarrow \norm[ij]{op} \visamp[ij]{o}\visamp[ij]{p},\\ 
  \triple[ijk]{opq} &\rightarrow \triple[ijk]{opq} 
                         \visamp[ij]{o}\visamp[jk]{p}\visamp[ki]{q}.
\end{align*}
with $\textvisamp[ij]{o}$ and $\textvisphi[ij]{o}$ the mean visibility amplitude and phase over the bandwidth.

 %%% CHROMATIC DISPERSION %%%

\subsection{Chromatic dispersion}
\label{ap:disp}

In this section we provide analytic formulas for the smearing of
interferometric observables when the quadratic dispersion term
$\textdisp[ij]{}$ cannot be neglected in the instrumental phase \insphi[ij]{}
(Eq.~\ref{eq:hyp:insphi}). It is only valid for the Gaussian
bandpass.  

In that case, two quadratic terms appear in the exponential of 
$\sflux[ij]{o}$ (Eq.~\ref{eq:cohernorm} with Eqs.~\ref{eq:hyp:insphi}
\& \ref{eq:hyp:bandpass} substituted in). One is the bandpass (real-valued)
and the other the dispersion (imaginary). They can be gathered by using a 
``complex'' resolution $\resol{}\rfact[ij]$ instead of $\resol{}$ in the
equations, where
\begin{equation}
  \rfact[ij] = \left( 1 - \j \frac{\disp[ij]{}}{4 \resol{}^2 \log 2} \right) ^ {\frac12}.
\end{equation}

From there calculations are very similar (involving affine transforms of
factor $\rfact[ij]$) and results of Sect.~\ref{sec:ana:clo} can be used provided
that the following substitutions are made:
\begin{subequations}
\begin{align}
  \phiobj[ij]{o}   &\rightarrow \phiobj[ij]{o} / \rfact[ij], \\
  \dopd[ij]        &\rightarrow \dopd[ij] / \rfact[ij], \\
  \phishift[ij]{}  &\rightarrow \phishift[ij]{} / \rfact[ij],
\end{align}
\end{subequations}
in Eqs.~(\ref{eq:gen:bispDS}, \ref{eq:gauss:bispDS} \& \ref{eq:bisp:gd}).

 %%% SMALL SMEARING %%%

\subsection{Approximation of small smearing}
\label{ap:smallsmearing}
For small enough baselines, the Taylor development of the complex exponential
in the  Fourier transforms can be used.  We find that the smearing
of the visibility can be linked to the moments of band pass
\begin{subequations}
\begin{align}
  \mom[ij]{op}(s) &= \intinf\sflux[i]{o}(\xi)\sflux[j]{p}(\xi)\xi^s\idiff\xi\\
\intertext{using}
  \decen[ij]{op} &= \frac{\mom[ij]{op}(1)}{\mom[ij]{op}(0)}
  \quad\text{(decentering)}
  \\
  \wid[ij]{op}
  &= 2\sqrt{\frac{\mom[ij]{op}(2)}{\mom[ij]{op}(0)}}
     \qquad\text{(width)},
    \\ 
  \skewness[ij]{op}
  &= \frac{\mom[ij]{op}(3)}{\mom[ij]{op}(0)}
     \qquad\text{(skewness)}.
\end{align} 
\end{subequations}

With these definitions the amplitude of the smearing is a second-order term of baseline while the phase-shift has first and third order terms:
\begin{subequations}
\begin{align}
  \envband[ij]{op}(\alpha)
  &= 1 - \frac{\wid[ij]{op}\!^2}{8\wavenumzero^2} \alpha^2 + O(\alpha^4),\\
  \phiband[ij]{op}(\alpha)
  &= \underbrace{\frac{\decen[ij]{op}}{\wavenumzero}}_{\approx 0} \alpha + \frac{\skewness[ij]{op}}{6\wavenumzero^3} \alpha^3 + O(\alpha^5).
\end{align} 
\end{subequations}
For sources with the same spectrum as the calibration lamp, $\textdecen[ij]{op}$ cancels out because of the spectral calibration. 

We conveniently chose to define the Gaussian-equivalent spectral resolution as 
\begin{equation}
  \resol{} = \frac{\wavenumzero}{2\sqrt{\log 2}\wid[ij]{op}},
\end{equation}
so that the formula of the Gaussian bandpass (Eq.~\ref{eq:gauss:vsqPS}) may be applied for small separations.

The Taylor developments of our two test-cases, top-hat and Gaussian bandpass,
are
\begin{subequations}
\begin{align}
  \textenvbandH[]{}(\alpha) &= 1 - (0.0416\dwavenum{}^2/\wavenumzero^2) \alpha^2 + O(\alpha^4),\\
  \textenvbandG[]{}(\alpha) &= 1 - (0.0451\dwavenum{}^2/\wavenumzero^2) \alpha^2 + O(\alpha^4).
\end{align}
\end{subequations}
This is not in agreement with \citet{ZHA07} who noticed a difference
by a factor of $1.5^2$ in the coefficient between both bandpasses. They may have assumed that the bandpass was given by the power spectral density
instead of the fringe envelope---there is a square involved, which leads to a factor 1.4 for the Gaussian FWHM. If we do the same, we then find results consistent with theirs.

In Figure~\ref{figap:accuracy} we show the error budget resulting from approximations in the treatment of the smearing.  While the source's spectrum
must be taken into account to achieve a good precision, we find that
in most cases a quadratic or exponential approximation for the smearing
amplitude $\Gamma$ is accurate enough.  The source's spectrum may be
implicitly taken into account by fitting the free parameter $\resol{}$
to the data.

\subsection{Smearing for the time-modulated ABCD method}
\label{ap:ABCD}

\begin{figure}
\includegraphics[width=\linewidth]{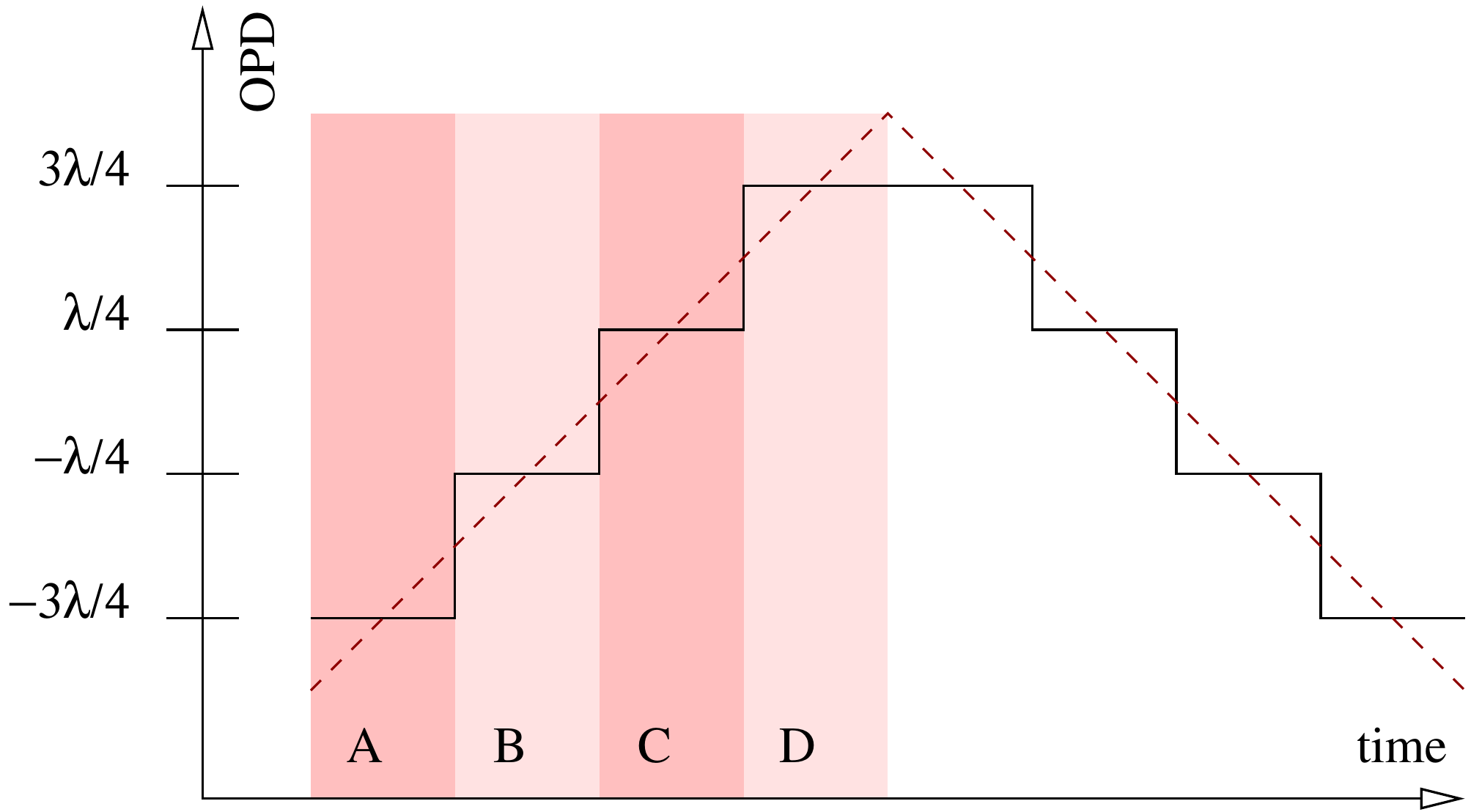}
\caption{Two scanning schemes for the ABCD method. \emph{Solid line:} by steps with measurements of the interferogram at OPDs $-3\lambda/4$, $-\lambda/4$, $\lambda/4$, $3\lambda/4$. \emph{Dashed line:} linearly with time with an averaging of the interferogram over OPD intervals of width $\pi/2$ centred on  $-3\lambda/4$, $-\lambda/4$, $\lambda/4$, $3\lambda/4$.}
\label{fig:ap:ABCD-scanning}
\end{figure}

The visibility amplitude and closure phase are deduced from the complex visibilities. For a white fringe centred at zero OPD and a sampling at OPD points $-3\lambda/4$, $-\lambda/4$, $\lambda/4$, $3\lambda/4$ for A, B, C, and D respectively, the non-normalised complex visibility is given by  
\begin{equation}
  \Nu^{ij} = [(B - D) + \j (A - C)] \exp{i\pi/4}.
\end{equation}
where $A$, $B$, $C$ and $D$ stand for the flux in the interferogram, i.e. the real part of $\textphasor[ij]{}(\delta)$.  We deal here with the OPD modulation that probes the interferogram at OPDs of exactly  $-3\lambda/4$, $-\lambda/4$, $\lambda/4$, $3\lambda/4$, which can be achieved in practice by a time-modulation of the OPD in steps (see Fig.~\ref{fig:ap:ABCD-scanning}).  For the other OPD modulation scheme, the variation is linear with time and the interferogram is integrated over an OPD interval of width $\pi/2$.  Our formulas are still a very good approximation for this scheme as long as there is some spectral resolution ($\resol{} \gtrsim 5$).

We substitute Eq.~\ref{eq:def:phasor} into A, B, C, and D to write
\begin{equation}
\begin{split}
  \Nu^{ij} &= 
    \sum_o \flux[ij]{o} \envband[ij]{o}(\phiobj{o} + \pi/4) 
      \cos\left(\phiobj{o} + \phiband[ij]{o}(\phiobj{o} + \pi/4)\right)\\
         &+ \j \sum_o \flux[ij]{o} \envband[ij]{o}(\phiobj{o} - \pi/4)
      \sin\left(\phiobj{o} + \phiband[ij]{o}(\phiobj{o} - \pi/4) \right),
\end{split}
\end{equation}
where the smearing terms read
\begin{align}
  \flux[ij]{o}             &= \big|\IFT{\sflux[ij]{o}}(0)\big|,\\
  \smearing[ij]{o}(\alpha) &= \frac{ \IFT{\sflux[ij]{o}}(\alpha - \pi/2) 
                                   + \IFT{\sflux[ij]{o}}(\alpha + \pi/2)}
                                   {\flux[ij]{o}},\\  
  \envband[ij]{o}(\alpha)  &= |\smearing[ij]{o}(\alpha)|, \\ 
  \phiband[ij]{o}(\alpha)  &= \arg \smearing[ij]{o}(\alpha). 
\end{align}

In the case of a linear instrumental phase (Eq.~\ref{eq:hyp:insphi} with no dispersion $\textdisp[ij]{}$) and top-hat or Gaussian transmission (assumed identical for all objects), the smearing amplitude is
\begin{subequations}
\begin{align}
  \envbandH{}(\alpha)     
  &=  \frac{
      \sinc\left(\frac{\alpha - \phishift[ij]{} - \frac\pi2}{2\resol{}}\right)
    + \sinc\left(\frac{\alpha - \phishift[ij]{} + \frac\pi2}{2\resol{}}\right)}{2},
  \\
  \envbandG{}(\alpha) 
  &= 
   \wideexp{\left(
     -\frac{(\alpha - \phishift[ij]{})^2 + \frac{\pi^2}4}{16\resol{}^2\log 2} 
   \right)}
     \cosh\left(\frac{\pi(\alpha - \phishift[ij]{})}{16\resol{}^2\log 2}\right),
  \label{eq:ABCD:gauss:smearing}
\end{align}
\label{eq:ABCD:beta}
\end{subequations}
the smearing phase reads
\begin{equation}
  \phiband{}(\alpha) = \insphi[ij]{}(0),
\end{equation}
and $\textflux[ij]{o}$ is proportional to the flux of the source.

Note that the visibility, in particular the amplitude, is impacted by the instrumental group delay $\rev{\textphishift[ij]{}}$ in contrary to the full scans processed with Fourier analysis.  The reason is that the ABCD only scans a small part of the interferogram \rev{where the smearing decreases the fringe contrast.  As a consequence, the visibility amplitude of spectrally dispersed fringes are biased by a chromatic instrument because the fringe-tracking is not able to zero the group delay in all channels simultaneously (see Appendix~\ref{ap:gd}).  Also, since the atmospheric differential piston introduces a random group delay (see Sect.~\ref{sec:atm:om}), the ABCD visibility amplitudes are sensitive to the atmospheric turbulence unless the fringe tracking is close to ideal. The workaround in both cases is to increase the spectral resolution so that the fringe contrast remains approximately constant over the OPD excursion due to the atmosphere and the instrumental group delay, typically several tens of microns.}

\beginrevision
\subsection{Group delay: fringe tracking \& atmospheric dispersion}
\label{ap:gd}

\begin{figure}
  \includegraphics[width=\linewidth]{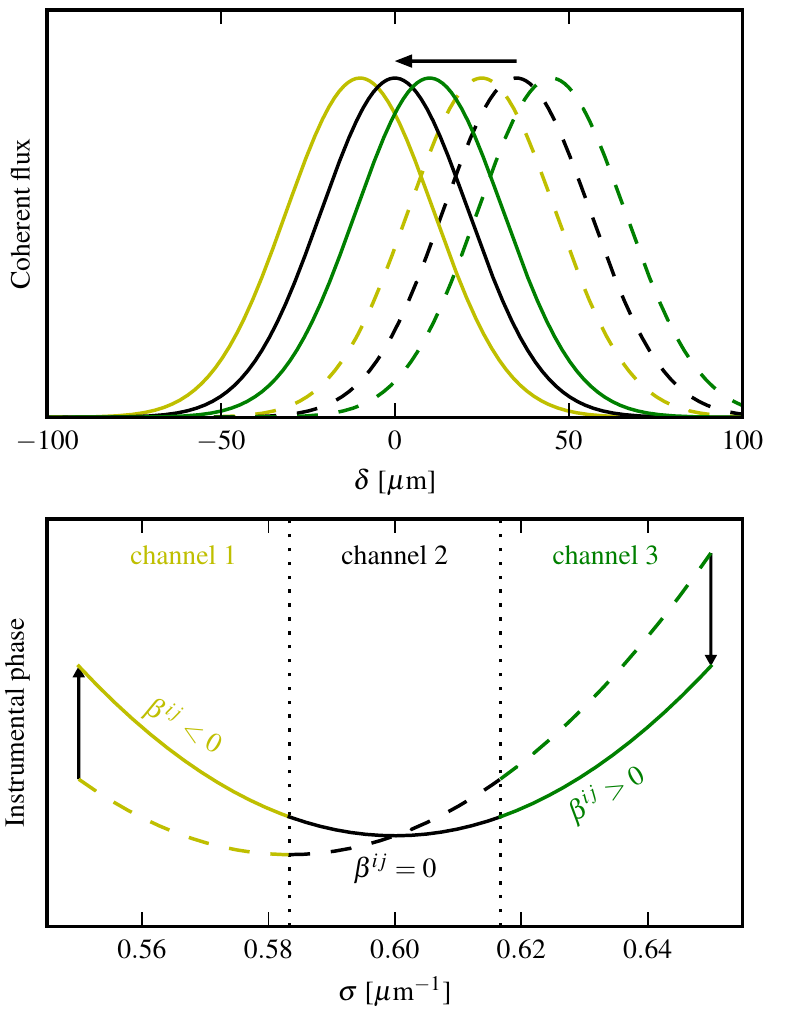}
  \caption{\rev{Fringe-tracking on the central spectral channel in a dispersive instrument with three spectral channels.  \emph{Top panel:} the position of the fringe packets before tracking is given by the dashed lines; after fringe-tracking, shown with an arrow, the central fringe packet is positioned at zero OPD (solid lines). \emph{Bottom panel:} the slope of the phase is modified by the tracking, making it zero on average on the central channel.  Because the phase is non linear, the average slope is non zero in the other channels, which explains why the corresponding fringe packets (top panel) are shifted with respect to zero OPD.} }
  \label{fig:atmgd}
\end{figure}

\begin{table}
  \caption{\rev{Group delay induced by the air dispersion in the delay lines of the VLTI, expressed per metre of interferometric delay.  The air dispersion has been taken from \citet{MAT07} with the typical temperature (17\textordmasculine C), humidity (16\%), and atmospheric pressure (743\,hPa) in the VLTI laboratory as reported by \citet{PUE06}.  Values at other sites may differ, but they have the same order of magnitude, except in the mid-IR ($LMNQ$) where humidity has a strong influence.}}
\label{tab:gdatm}
\begin{tabular}{lccc}
\hline\hline
band & wavelength & max. group delay & max. fringe packet shift\\
     &            & $\phishift[ij]{}$ & $\phishift[ij]{}/2\pi\wavenumzero$\\
     & [$\mu$m]   & [rad/m] & [$\mu$m/m]\\
\hline
$J$  & 1.12--1.32 & $2\pi\times 0.340$   &0.38\\
$H$  & 1.48--1.78 & $2\pi\times 0.157$   &0.25\\
$K$  & 2.00--2.38 & $2\pi\times 0.040$   &0.08\\
$L$  & 3.22--3.68 & $2\pi\times 0.071$   &0.25\\
$M$  & 4.52--4.98 & $2\pi\times 0.115$   &0.52\\
$N$  & \hphantom{0}8.2--12.2  
                  & $2\pi\times 0.014$   &0.17\\
$Q$  & 18.5--23.5 & $2\pi\times 0.044$   &0.82\\
\hline
\end{tabular}
\end{table}

We consider a state of the interferometer, atmosphere, and recombiner in which the ``instrumental'' differential phase is $\insphi[ij]{*}(\wavenum)$.  If the fringe-tracker introduces an OPD $\opdvar$, the instrumental phase 
becomes
\begin{equation}
  \insphi[ij]{}(\wavenum) = \insphi[ij]{*}(\wavenum) - 2\pi\wavenum\opdvar.
  \label{eq:5:1}
\end{equation}
In group-delay tracking around wavenumber $\wavenumzero$ 
\begin{equation}
  \delta = \frac{1}{2\pi} \pderiv{\insphi[ij]{*}}{\wavenum}(\wavenumzero) 
  \label{eq:5:2}
\end{equation}
so that, in the end, the group delay at reference wavenumber 
\begin{equation}
  \phishift[ij]{0}
  = \wavenumzero \pderiv{\insphi[ij]{}}{\wavenum}(\wavenumzero) 
\end{equation}
becomes zero.  If several channels are present, the group delay in channel $c$ with central wavenumber $\wavenum_c$ is
\begin{equation}
\label{eq:5:4}
\begin{split}
  \phishift[ij]{c}
  &= 
    \wavenum_c \pderiv{\insphi[ij]{}}{\wavenum}(\wavenum_c), 
  \\
  &\approx
    \wavenum_c (\wavenum_c - \wavenumzero)
    \pderiv[2]{\insphi[ij]{*}}{\wavenum}(\wavenumzero).
\end{split}
\end{equation}
Figure~\ref{fig:atmgd} shows how fringe tracking can cancel the group delay (and instrumental phase slope) at a reference channel but not at adjacent ones.  This is due to the non linearity of the instrumental phase as the second derivative in Eq.~(\ref{eq:5:4}) shows.

In the particular case of differential air dispersion in the delay lines, the
additional differential phase is
\begin{equation}
  \insphi[ij]{\text{atm}}(\wavenum) = 2\pi\textdelay[ij] \wavenum r(\wavenum)
  \label{eq:5:3}
\end{equation}
where \textdelay[ij] is the interferometric delay and $r(\wavenum)$ is the refractive index of air minus 1 at wavenumber $\wavenum$. Using Eqs.~(\ref{eq:5:1}, \ref{eq:5:2}) we write interferometer transmission as
\begin{equation}
\begin{split}
  \strans[ij]{\text{int}}(\sigma) &= 
    \text{const.} \times 
    \wideexp{
       \j \left[
          \insphi[ij]{\text{atm}}(\wavenum) 
        - \wavenum
        \pderiv{\insphi[ij]{\text{atm}}}{\wavenum}(\wavenumzero)
          \right] 
      },\\
      &\approx 
      \text{const.}^{\prime} \times \wideexp{
      \left[\frac\j2 (\wavenum-\wavenumzero)^2
    \pderiv[2]{\insphi[ij]{\text{atm}}}{\wavenum}(\wavenumzero) \right]}
\end{split}
\end{equation}
The corresponding group delay (Eq.~\ref{eq:5:4}) has been computed for the different infrared bands and its typical value is reported in Table~\ref{tab:gdatm}. In the $H$ band, the atmospheric group delay becomes of the order of PIONIER's internal group delay ($\sim 10\,\mu$m, see Fig.~\ref{fig:PT}) for interferometric delays of about 50 metres. Near-zenith observations with 100~metres baselines are therefore more impacted by PIONIER's instrumental phase. 
\endrevision

\section{Validity of the Gaussian approximation}
\label{ap:closim}

\begin{figure*}
  \centering
  \includegraphics[width=.8\linewidth]{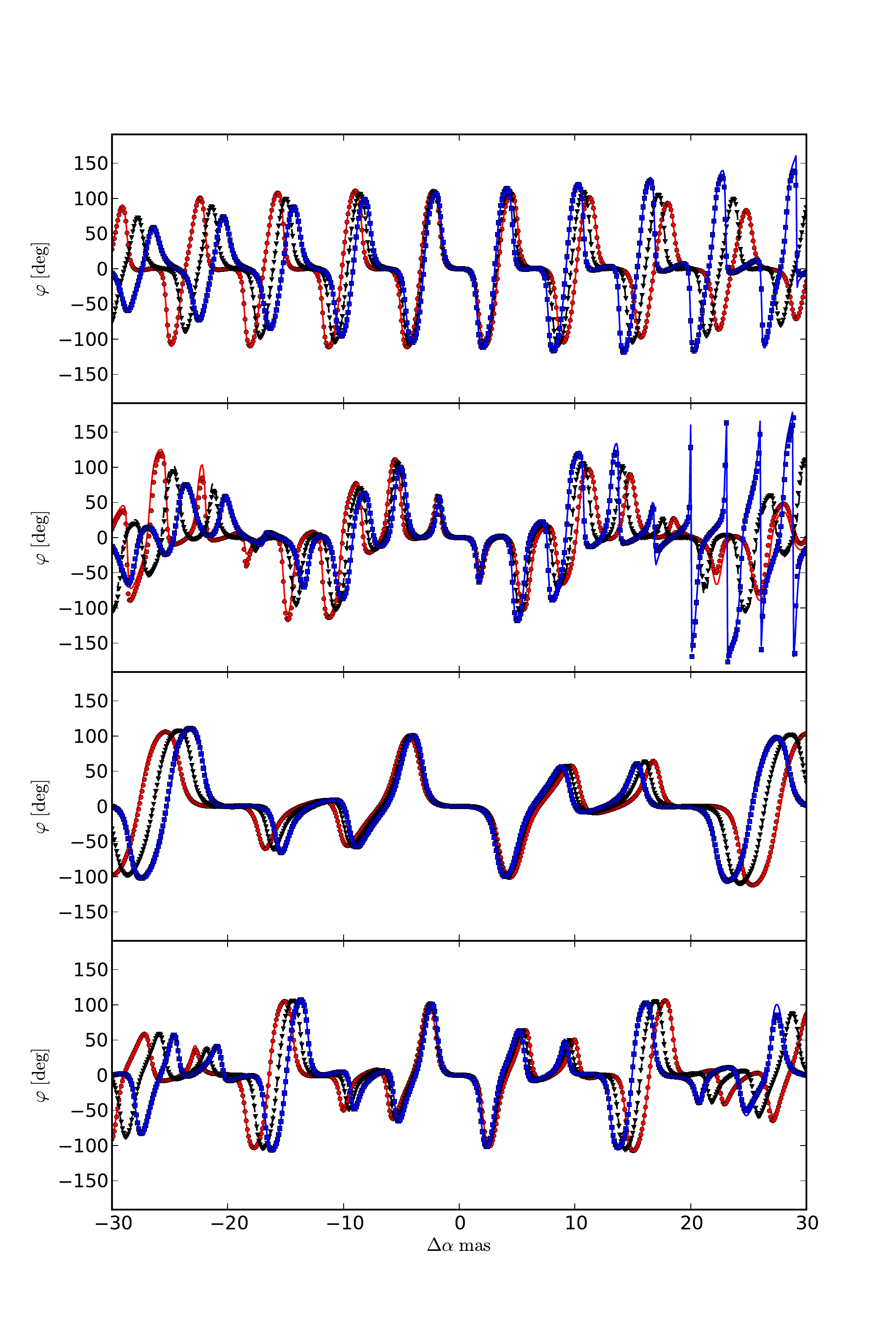}
  \caption{Closure phase of PIONIER on the four telescope triplets using the 
    baseline configuration of Fig.~\ref{fig:uv}. A 
    binary of flux ratio 0.6 with separations \rev{from} $-30$ to 30\,mas in the 
    East-West direction has been used.  The three pairs of markers and line 
    correspond to the three spectral channels in the H band.
    \emph{Ma\rev{r}kers:} closure phases obtained from
  simulated fringes \emph{Lines:} analytic model of Sect~\ref{sec:ana}.} 
  \label{fig:closim}
\end{figure*}

When the amount of smearing becomes significant, the Gaussian model and actual smearing differ significantly.  Figure~\ref{fig:closim} gives a comparison between the simulated and modelled closure as a function of separation; the model assumes a binary with a West-East separation observed in the test PIONIER configuration of Sect.~\ref{sec:bin}.  It is worth noting that the antisymmetry of the closure phase (with respect to orientation of the binary) is broken by the instrumental signatures; also the different spectral channels display different behaviours. 

\end{document}